\documentclass[twocolumn,usenatbib]{mnras}

\usepackage{subcaption}
\usepackage{graphicx}
\usepackage{amsmath}
\usepackage{siunitx}
\usepackage{physics}
\usepackage{xfrac}
\usepackage{comment}
\usepackage{hyperref}

\def\pmb#1{\setbox0=\hbox{#1}%
    \kern-.025em\copy0\kern-\wd0
    \kern.05em\copy0\kern-\wd0
    \kern-.025em\raise.0433em\box0}

\def\ltsima{$\; \buildrel < \over \sim \;$}
\def\gtsima{$\; \buildrel > \over \sim \;$}
\def\simlt{\lower.5ex\hbox{\ltsima}}
\def\simgt{\lower.5ex\hbox{\gtsima}}
\def\p2Y{\;_2Y}
\def\m2Y{\;_{-2}Y}

\def\mk2{\mu {\rm K}^2}
\def\Planck{\it Planck \rm}

\def\Planckns{\it Planck\rm}
\def\LCDM{$\Lambda{\rm CDM}$ }
\def\AE{\textit{Paper I}}
\def\LCDMns{$\Lambda{\rm CDM}$}

\def\pmb#1{\setbox0=\hbox{#1}%
     \kern-.025em\copy0\kern-\wd0
     \kern.05em\copy0\kern-\wd0
     \kern-.025em\raise.0433em\box0}

\usepackage{xcolor}

\definecolor{purple}{RGB}{156,81,182}

\begin{document}

\title[$S_8$ tension II]{A non-linear solution to the $S_8$ tension II: Analysis of  DES Year 3 cosmic shear}

\author[Preston, Amon \& Efstathiou]{Calvin Preston$^{1,2}$\thanks{E-mail: cp662@cam.ac.uk}, Alexandra Amon $^{1,2}$\thanks{E-mail: alexandra.amon@ast.cam.ac.uk}, George Efstathiou $^{1,2}$\thanks{E-mail: gpe@ast.cam.ac.uk} \\
${1}$ Kavli Institute for Cosmology Cambridge,
Madingley Road, Cambridge, CB3 OHA. \\
${2}$ Institute of Astronomy. Madingley Road, Cambridge, CB3 OHA. United Kingdom.}

\maketitle

\begin{abstract} 
Weak galaxy lensing surveys have consistently reported low values of the $S_8$ parameter compared to the \Planckns\ \LCDMns\ cosmology. \citet{AAGPE2022} used KiDS-1000 cosmic shear measurements to propose that this tension can be reconciled if the matter fluctuation spectrum is suppressed more strongly on non-linear scales than assumed in state-of-the-art hydrodynamical simulations. In this paper, we investigate cosmic shear data from the Dark Energy Survey (DES) Year 3. The non-linear suppression of the matter power spectrum required to resolve the $S_8$ tension between DES and the \Planckns\ \LCDMns\ model is not as strong as inferred using KiDS data, but is still more extreme than predictions from recent numerical simulations. An alternative possibility is that non-standard dark matter contributes to the required suppression. We investigate the redshift and scale dependence of the suppression of the matter power spectrum. If our proposed explanation of the $S_8$ tension is correct, the required suppression must extend into the mildly non-linear regime to wavenumbers $k\sim 0.2 h {\rm Mpc}^{-1}$. In addition, all measures of $S_8$ using linear scales should agree with the \Planckns\ \LCDMns\ cosmology, an expectation that will be testable to high precision in the near future.
\end{abstract}

\begin{keywords}
cosmology: cosmological parameters, weak lensing, observations,
\end{keywords}

\section{Introduction}\label{sec:intro}

The six-parameter \LCDM cosmological model has been incredibly successful in explaining the anisotropies of the cosmic microwave background \citep[CMB; e.g.][]{Params:2018}, baryon acoustic oscillations \citep[BAO; e.g.][] {Alam:2021a} and a wide range of other astronomical data. Nevertheless, there are indications of 'tensions'  between this model and some observations. The `Hubble tension', i.e. the discrepancy between distance ladder measurements of the Hubble parameter at present day, $H_0$, and the value inferred from the CMB assuming the \LCDM model, is the most well known \citep[for reviews see][] {Freedman:2021, Shah:2021, Kamionkowski:2022}. In addition, there is an apparent discrepancy between the amplitude of the matter fluctuations inferred from the CMB and that measured in cosmic shear surveys which has become known as the `$S_8$ tension'\footnote{Where $S_8 = \sigma_8 (\Omega_{\rm m}/0.3)^{0.5}$, $\Omega_{\rm m}$ is the present day matter density in units of the critical density and $\sigma_8$ is the root mean square linear amplitude of the matter fluctuation spectrum.}, \citep[e.g.][]{Heymans:2013, Asgari:2021, Amon:2021, Secco:2022, dalal2023hyper, li2023hyper}. Since new physics may be required to explain these tensions, they have become the focus of many recent observational and theoretical studies.  

This paper is a sequel to \citet{AAGPE2022}, hereafter \AE, and is devoted exclusively to the $S_8$ tension. \AE\ investigated the hypothesis that the \Planck \LCDM cosmology accurately describes matter fluctuations on linear scales, including their growth rate. The $S_8$ tension is then explained by modifying the matter power spectrum on non-linear scales. \AE\ adopted a simple phenomenological model for the matter power spectrum
\begin{equation}
P_{\rm m}(k, z) =  P^{\rm L}_{\rm m}(k, z) + A_{\rm mod}[P^{\rm NL}_{\rm m} (k, z) - P^{\rm L}_{\rm m}(k, z)] \,, \label{equ:NL}
\end{equation} 
where the superscript ${\rm L}$ denotes the linear theory power spectrum and the superscript NL denotes the non-linear power spectrum in a model in which the matter behaves like cold dark matter (i.e. ignoring the thermal pressure of baryons and  baryonic feedback). The $A_{\rm mod}$ parameter modulates the amplitude of the non-linear spectrum and  can describe a suppression of power on small scales. \AE\ compared this model to the weak lensing measurements from the Kilo-Degree Survey (KiDS) reported by \cite{Asgari:2021} and showed that the \Planck\ \LCDM cosmology provides acceptable fits to the KiDS shear-shear two-point statistics if the suppression parameter has a value $A_{\rm mod} \approx 0.69$. Our hypothesis explains why the \Planck\ \LCDM\ cosmology agrees so well with: (a) the background expansion history measured from Type Ia supernovae over the entire redshift range $0.1 \simlt z \simlt 1.5$ spanned by the Pantheon sample, \citep[e.g.][]{Brout:2021} and (b) gravitational lensing of the CMB \citep{Plensing:2020, Darwish:2021, Omori:2022}, since these measurements are dominated by linear scales and are consistent with the predictions of General Relativity (i.e. there is  no evidence for `gravitational slip' as might be expected in theories of modified gravity, see e.g. \citealt[][]{Bertschinger:2011}). In fact, point (b) has been strongly reinforced with the recently released CMB weak lensing results from the Atacama Cosmology Telescope (ACT), which are in excellent agreement with the \Planck\ \LCDM\ cosmology \citep[][and references therein]{ACT_madhavacheril2023}.

Our hypothesis predicts that {\it all} measures of the fluctuation amplitude dominated by linear scales, such as galaxy redshift-space distortions \citep[e.g.][]{Alam:2021a, Philcox:2022, Chen:2022, DAmico:2022} and cross-correlations of CMB lensing with galaxy surveys \citep[e.g.][]{Chang:2022, White2022, ChenWhite:2022} should agree with the predictions of the \Planck \LCDM model. As discussed in \AE\, there is not yet a consensus on the interpretation of these types of measurements, with some authors reporting tension with \LCDMns. This situation is likely to change in the near future, for example, via cross-correlations of the new ACT lensing maps and galaxy surveys and redshift-space distortion measurements with the Dark Energy Spectroscopic Instrument, \cite[DESI]{DESI:2016}. Should such measurements conflict with \LCDM\ on linear scales, our proposed solution of the $S_8$ tension will become untenable. 

If the hypothesis of \AE\ is correct, what physical processes might be responsible for a suppression of power on small scales? Baryonic feedback is an obvious candidate, though the suppression required to explain the KiDS data is stronger than seen in recent cosmological hydrodynamical simulations \citep[e.g.][]{Dubois:2014, McCarthy:2017, Springel:2018}. The physics of feedback is, however, complex and is not yet sufficiently well understood to exclude baryonic feedback as the sole cause of the suppression. Suppression of the power spectrum can also be achieved by invoking  more complex  models of dark matter, for example, adding a component of warm or axionic dark matter \citep[see e.g.][]{Widrow:1993, Hu:2000, Hui:2017, Rogers:2023} to the cold dark matter of \LCDMns. Since it may be difficult to disentangle the effects of baryonic feedback from those of exotic dark matter, we adopt a phenomenological approach to modelling the power spectrum on non-linear scales. An alternative approach, based on ad hoc modifications to the halo mass function, has been described recently by \cite{Gu:2023}. As in \AE, we remain agnostic as to the exact physical cause of power spectrum suppression.

In this paper, we analyse the cosmic shear two-point statistics from the Dark Energy Survey (DES) Year 3 analysis presented in \citet{amon:2022}; \citet*{Secco:2022}, hereafter DES22. The DES measurements are mostly independent of those from KIDS and have somewhat higher statistical power. Our analysis of DES, therefore, provides a check of the results of \AE. In this work, we extend the analysis of \AE\ by generalising  Eq.~\ref{equ:NL} to allow the power suppression to vary with redshift and wavenumber. As we will see it is difficult to extract detailed information on these dependencies with the current generation of weak lensing surveys.

This paper is structured as follows. Sec.~\ref{sec:des} discusses the Dark Energy Survey, the data used and the modelling choices made in our analysis. We compare the DES and KiDS constraints on the parameter  $A_{\rm mod}$ in Sec.~\ref{sec:Amod}. Sec.~\ref{sec:Cmodz} explores constraints imposed by the DES measurements on the redshift dependence of the power suppression. In Sec.~\ref{sec:abin}, we explore the scale dependence of the suppression to provide more stringent tests of whether baryon feedback can resolve the $S_8$ tension. Finally, we summarize our conclusions and discuss the implications of our results in Sec.~\ref{sec:discussion}.

\section{Dark Energy Survey}
\label{sec:des}

\begin{figure*}
\centering
\includegraphics[width=\textwidth]{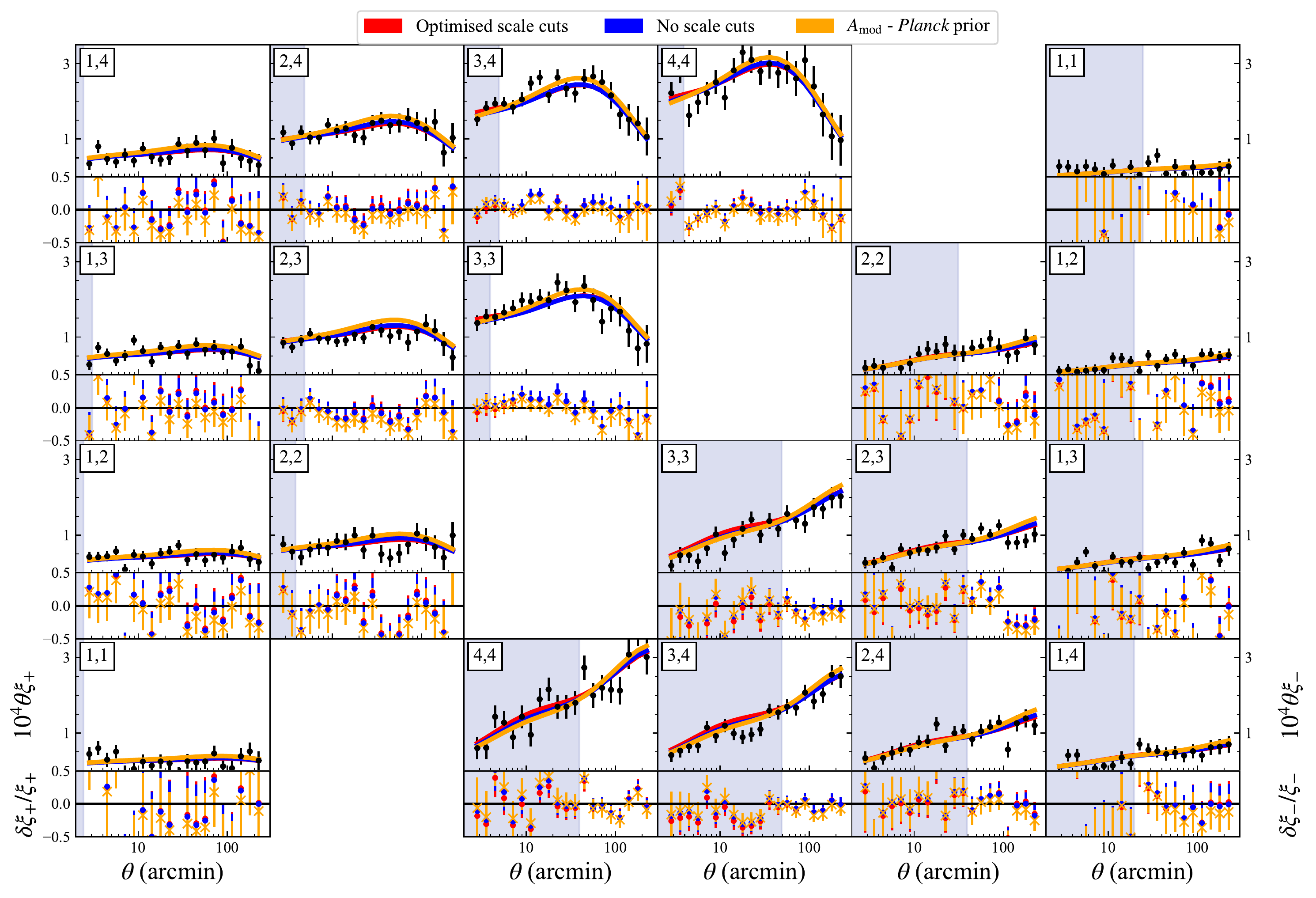} 
    \caption{The DES Y3 $\xi_{\pm}$ cosmic shear measurements of \citet{amon:2022}; \citet*{Secco:2022} and best-fit theoretical models using three modelling choices. In red we show the best fit of variant 3: when `\LCDMns-Optimised' scale cuts are applied to mitigate the effects of baryonic feedback, following DES22
    %, and with {\sc HMCode2020} as the nonlinear model, including no baryonic feedback parameter. 
    The blue line shows the best fit if all scales are used and baryonic feedback is ignored. The yellow line denotes the best fit when all scales are used, the $A_{\rm mod}$ parameter is included and the cosmology is based on the Planck prior as described in Sec.~\ref{sec:des}. All three provide good fits, with very similar values for the reduced-$\chi^{2}$, quoted in Table~\ref{tab:chi2}. Residuals are shown below each panel, offset from one another to be visible.}
	\label{fig:OptSCvsNoSCvsAmodDV}
\end{figure*}

The Dark Energy Survey \citep[DES][]{DES:2018, DES:2021} is a six-year imaging programme that observed $\sim5000\ $deg$^{2}$ of the Southern Hemisphere in five photometric bands. In this paper, we use data from the first three years of observations (DES Y3) which span the complete footprint of the full survey  at a reduced depth. The DES Y3 cosmic shear two-point correlation function measurements presented in DES22 are based upon the shape estimations of over 100 million galaxies \citep{Gatti:2021}, calibrated using a suite of image simulations \citep{MacCrann:2022}. The data is divided into four redshift bins and the redshift distributions are calibrated to give an overall mean redshift of $z = 0.63$ \citep{Myles:2021}.

We reanalyse the DES cosmic shear data using the public DES pipeline and following the analysis of DES22\footnote{Following the DES Y3 analysis, we use nested sampling via the \textsc{Cosmosis} cosmology parameter estimation framework \citep[][]{Zuntz:2015}, but instead use the \textsc{MultiNest} sampler \citep{Feroz_2009}, as used in \AE. For this specific analysis, investigations of a subset of models made with the sampler \textsc{Polychord} \citep{Handley:2015} gave consistent results, though requiring substantially more computing time for convergence.}. We update the analysis choices as follows: we assume three neutrino species with two massless states and one massive state with a mass of $0.06 \ {\rm eV}$; we adopt the non-linear alignment (NLA) model for intrinsic alignments as described in \cite*{Secco:2022}; we use {\sc HMCode2020} \citep{Mead:2021} to  model the nonlinear matter power spectrum, which generates a non-negligible shift in the $S_8$ posterior compared to the {\sc Halofit} model \citep{Takahashi:2012} used in DES22. These analysis choices are tested and compared in detail in \cite{KiDSDES}. 

Table~\ref{tab:priors} lists the priors used in this paper. For the `Free' cases, we applied uninformative priors on the cosmological parameters, following DES22, with the exception of the neutrino mass which is kept fixed in our analysis. The remaining entries in Table~\ref{tab:priors} give the priors for the redshift calibration for each bin, $\Delta z$, the shear calibration for each bin, $m$, and the single intrinsic alignment parameter $A_1$ of the NLA model, as in DES22.

In \AE, which was exploratory in nature, we demonstrated that the best fit \Planckns\ \LCDMns\ cosmology provides acceptable fits to the KiDS shear-shear statistics if $A_{\rm mod}\sim 0.7$. However, we did not account for any uncertainty on the cosmological parameters, particularly $S_8$, as measured by \Planckns. As a consequence ,\AE\ overestimated the suppression of small scale power required to resolve the $S_8$ tension with KiDS. In this paper, instead of a joint analysis, we include the \Planckns\ uncertainty on $S_8$ by applying a prior on $S_8$ and $\Omega_{\rm m}$ derived from the base \LCDMns\ \Planckns\ TTTEEE chains:

\begin{eqnarray}
\chi^2_{\rm Planck} &=&   19249.9592 (S_8 - 0.8275)^2    \nonumber \\
 &-& 65748.1012(S_8-0.8275)(\Omega_{\rm m}-0.3134)  \nonumber  \\
 &+&  71122.5486(\Omega_{\rm m}-0.3134)^2.  \label{equ:Planckprior}  
\end{eqnarray}
This prior is described in more detail in Appendix~\ref{sec:planckprior}. The scalar spectral index and physical baryon density are poorly determined by DES and KiDS but are determined to high-precision by \Planck\ within the context of the six-parameter \LCDM\ cosmology. Therefore, we fix these parameters to the best-fit \Planck\ values from \citet{EfstathiouGratton:2021}: $n_{\rm s} = 0.9671$ and $\omega_{\rm b} = \Omega_{\rm b} h^2 = 0.02226$. Finally, for a given value of $\Omega_{\rm m}$, the Hubble parameter $h$ is set from the well-determined parameter combination  $\Omega_{\rm m}h^3 = 0.09612$. Applying a prior in this way allows us to accurately account for the \Planck\ uncertainty without needing to evaluate the \Planck\ likelihood in our analysis.

\begin{table}
    \caption{Summary of parameters and their priors used in the analysis. The prior ranges for parameters with flat priors and given in square brackets with the prefix F. The mean and $1\sigma$ width for parameters with Gaussian priors are given by the brackets with prefix $G$. The prior ranges listed here are those used in the `free' cosmology analysis throughout this paper. The third column lists the fiducial \Planck TTTEEE \LCDM parameters for the 10.5HM likelihood as given in \citet{EfstathiouGratton:2021}. In this paper, when we apply a \Planckns\ prior,  we fix the cosmological parameters that are unconstrained by weak lensing to the fiducial \Planckns\ values, but we vary $\Omega_{\rm m}$ and $S_8$ using the prior defined in Eq.~\ref{equ:Planckprior}. For intrinsic alignments, we chose to use the non-linear alignment model (NLA) for this analysis using an uninformative prior for the tidal alignment parameter $a_{1}$. We use the data calibration parameters recommended in the fiducial DES analysis \citet{DES:2021}. When using the \Planck prior, the values of the nuisance parameters are fixed to their best-fit value from the DESY3 analysis.}
\label{tab:priors}
\begin{center}
\begin{tabular}{lll}
Parameter & `Free' prior & `\Planck' \tabularnewline
%Parameter & prior & prior \tabularnewline
\hline 
\bf{Cosmological} \tabularnewline
$\Omega_{\rm m}$ \  Total matter density & F$[0.1, 0.9]$ & - \tabularnewline
$\Omega_{\rm b}$  \  Baryon density & F$[0.03, 0.07]$ & - \tabularnewline

$\Omega_{\rm b}h^{2}$   & - & 0.02226 \tabularnewline
%Physical baryon density
$10^{-9}A_{\rm s}$ Scalar spectrum amp. & F$[0.5, 5.0]$ & - \tabularnewline
$h $  \  Hubble parameter & F$[0.55, 0.91]$ & -  \tabularnewline
$\Omega_{\rm m}h^{3}$  \  & - & 0.09612 \tabularnewline
$n_{\rm s}$ \ Spectral index & F$[0.87,1.07]$ & 0.9671 \tabularnewline
$m_\nu$ \ Neutrino mass $({\rm eV})$  & $0.06 $ & 0.06 \tabularnewline
\hline 
\bf{Systematic} \tabularnewline
$A_1$ \ \ Intrinsic alignment amp.  & F$[-5,5]$ & 0.131  \tabularnewline
$\Delta z^1$ \  Source redshift 1 & $G( 0.0, 0.018 )$ & 0.004  \tabularnewline
$\Delta z^2$ \ Source redshift 2 & $G( 0.0, 0.015 )$ & -0.001 \tabularnewline
$\Delta z^3$ \ Source redshift 3 & $G( 0.0, 0.011 )$ & 0.001 \tabularnewline
$\Delta z^4$ \ Source redshift 4 & $G( 0.0, 0.017 )$ & 0.001 \tabularnewline
$m^1$ \ \ Shear calibration 1 & $G( -0.006, 0.009 )$ & -0.006\tabularnewline
$m^2$ \ \ Shear calibration 2 & $G( -0.020, 0.008 )$ & -0.019\tabularnewline
$m^3$ \ \ Shear calibration 3 & $G( -0.024, 0.008 )$ & -0.025\tabularnewline
$m^4$ \ \ Shear calibration 4 & $G( -0.037, 0.008 )$ & -0.037\tabularnewline

\hline 

\end{tabular}
\end{center}
\end{table}

Results of analysing the DES Y3 $\xi_{\pm}$ data for the different parameterisations and modelling choices discussed in this paper are summarized in Table~\ref{tab:chi2}. The first two entries (labelled 1 and 2), summarize results from DES22 (see their Table III) using their `\LCDM\-Optimised' scale cuts, which excludes data points lying within the shaded regions in Fig.~\ref{fig:OptSCvsNoSCvsAmodDV} from the analysis. Both of these analysis variants use {\sc Halofit} to model the non-linear power spectrum. Variant 1 uses the 5-parameter Tidal Alignment and Tidal Torquing (TATT) model for intrinsic alignments  and allows the neutrino mass to vary, as described in DES22. In variant 2, the neutrino mass is fixed to $0.06\ {\rm eV}$ and the two-parameter NLA model, a sub-space of the TATT model which includes a redshift dependence, is used to model intrinsic alignments, resulting in a $\sim1\sigma$ shift toward a higher value of $S_8$.

\setlength\extrarowheight{4pt}
\begin{table*}
  \centering
  \caption{Mean posterior value of $S_8$ and 1$\sigma$ error. $\chi^2_{\rm min}$ gives the minimum values of  $\chi^2$ for each analysis variant. Here, $\hat\chi^2_{\rm min}=\chi^2_{\rm min}/N_{\rm deg}$, where $N_{\rm deg}$ is the number of degrees of freedom. Following \AE, for a number of data points $N_{\rm DP}$, we have $N_{\rm deg}=N_{\rm DP}-4.5$ for free fits and $N_{\rm deg}=N_{\rm DP}-2.5$ for \Planck fits \citep{AAGPE2022}. This counting of effective degrees of freedom is approximate but relatively unimportant since $N_{\rm DP}$ is much greater than unity. The column labelled $N_{\sigma}$ lists $(\chi^{2}-N_{\rm deg})/\sqrt{2N_{\rm deg}}$. The column labelled $N_{\rm FP}$ lists the number of free parameters in each model. The starred indexes report two results from DES22 (see their Table 3, entries 2 and 3). The first is the \LCDM\-Optimised result, ie applying scale cuts (SC). The second uses the simpler NLA model for intrinsic alignments and fixes the neutrino mass. These two changes in analysis choices are maintained throughout Variants 3-10. Variant 3 (and subsequent variants) uses {\sc HMCode2020} to model the non-linear matter power spectrum without feedback. $A_i$ extensions sample the power spectrum in $5$ bins in wavenumber $k$ (Sec.~\ref{sec:abin}). Variants 12-13 report the constraints on $S_8$ using data from KiDS. All other variants use DES Y3. Cases labelled `\Planck' indicate those where the \Planck prior is applied to the cosmological parameters, rather than uninformative priors, as in the `free' case. }
\begin{tabular}{ccccccccccc}
\hline
Variant & Data & Small-scale model & Other  & Cosmology & $S_8$ & $N_{\rm DP}$ & $N_{\rm FP}$ & $\chi^2_{\rm min}$ & $N_\sigma$ &$\hat\chi^2_{\rm min}$  \tabularnewline
\hline
         1* & DES & SC & {\sc Halofit}, TATT, free-$\nu$  & Free & $0.772^{+0.018}_{+0.017}$ & 273 & 14 &  285.0 & 1.14 & 1.06  \tabularnewline
         2* & DES& SC & {\sc Halofit} & Free & $0.788^{+0.017}_{+0.016}$ & 273 & 14 &  288.1 & 1.28 & 1.07  \tabularnewline
\hline   
        3 & DES& SC&  & Free & $0.805\pm{0.017}$ & 273 & 14 & 283.7 & 0.66 & 1.06  \tabularnewline
        4 & DES& - &  & Free & $0.793\pm{0.012}$ & 400 & 14 & 418.1 & 0.80 & 1.06  \tabularnewline
        5 & DES& $A_{\rm mod}$ & & Free & $0.831\pm{0.036}$ & 400 & 15 & 417.1 & 0.77 & 1.05  \tabularnewline
        6 & DES&  $A_{\rm mod}$ & & \Planck  & $0.811\pm{0.010}$ & 400 & 3 & 420.7 & 0.80 & 1.06  \tabularnewline
        7 & DES&  $A_{\rm mod}$ + SC &  & \Planck & $0.807\pm{0.012}$ & 273 & 3 & 289.4 & 0.81 & 1.07   \tabularnewline
        8 & DES& $C_{\rm mod}(z)$ & & Free & $0.842\pm{0.022}$ & 400 & 20 & 413.1 & 0.63 & 1.04  \tabularnewline
        9 & DES& $C_{\rm mod}(z)$ & & \Planck & $0.813\pm{0.009}$ & 400 & 8 & 419.3 & 0.77 & 1.05  \tabularnewline
        10 & DES& $A_i$   & & Free & $0.810\pm{0.025}$ & 400 & 19 & 414.2 & 0.85 & 1.05 \tabularnewline
        11 & DES& $A_i$ & & \Planck   & $0.813\pm{0.009}$ & 400 & 7 & 418.1 & 0.73 & 1.05  \tabularnewline
        \hline
        12 & KiDS &  $A_{\rm mod}$ & & Free  & $0.780\pm{0.035}$ & 225 & 21 & 260.3 & 1.89 & 1.18 \tabularnewline
        13 & KiDS &  $A_{\rm mod}$& & \Planck  & $0.808\pm{0.013}$ & 225 & 3 &  265.5 & 2.04 & 1.19  \tabularnewline
        \hline
\end{tabular}
\label{tab:chi2}
\end{table*}

Variant 3 in Table~\ref{tab:chi2} uses {\sc HMCode2020} dark matter model for the non-linear power spectrum, instead of {\sc Halofit} and simplifies the intrinsic alignment model even further using the NLA model without a redshift-dependence (which \citet*{Secco:2022} have shown gives an acceptable fit to the DES Y3 data). As in variant 2, the cosmological and nuisance parameters are allowed to vary over the prior ranges given in Table~\ref{tab:priors}. With this set of analysis choices, the mean posterior value of $S_8$ in variant 3 is about $1\sigma$ higher than the value from variant 2. We refer the reader to \cite{KiDSDES} for a detailed analysis of the impact of each these modelling strategies.
%, with the difference almost entirely attributed to using {\sc HMCode2020} to model the non-linear power spectrum \cite{KiDSDES}. Furthermore, \cite{KiDSDES} show that {\sc HMCode2020} is more accurate than {\sc Halofit} for \LCDM\ like models.  

Variant 4 is the same as variant 3 except that it uses $\xi_\pm$ measurements over the entire angular ranges plotted in Fig.~\ref{fig:OptSCvsNoSCvsAmodDV} \footnote{Note that although the DES Y3 scale cuts were chosen to reduce the sensitivity of $S_8$ to baryonic feedback, small angular scales are subject to other systematics. The most important of these concerns is the modelling of the point spread function, which has been shown to be unimportant on the angular scales used in DES22. Further work is required to assess the impact of PSF uncertainties on $\xi_\pm$ on the angular scales falling within the shaded regions of Fig.~\ref{fig:OptSCvsNoSCvsAmodDV}.}. The best-fit models from variants 3 and 4 are compared to the DES Y3 measurements in Fig.~\ref{fig:OptSCvsNoSCvsAmodDV} and although baryon feedback is not accounted for in variant 4, they both provide excellent (almost indistinguishable) fits to the data. The $\chi^{2}$ values for these fits are given in Table~\ref{tab:chi2}.

Comparing the posterior widths of $S_8$ with the \Planck\ TTTEEE result of $S_8 = 0.828 \pm 0.016$ \citep{EfstathiouGratton:2021}, we see that in variants 3 and 4, $S_8$ is lower by about $0.9\sigma$ and $1.8 \sigma$, respectively. One might therefore conclude that with the choices made in this paper (i.e. adopting  the NLA intrinsic alignment model and HMCode2020) the DES Y3 weak lensing measurements are consistent with \Planck\ \LCDM {\it even if baryonic feedback effects are ignored}. This inference holds, apparently, even for variant 4, which  has a smaller error on $S_8$ and is sensitive to spatial scales at which all cosmological hydrodynamical simulations show power spectrum suppression caused by baryonic feedback.  However, focusing on the $S_8$ parameter gives a misleading impression of consistency because cosmological parameters allowed by DES are disfavoured by \Planckns. As we will show in the next section, suppression of the non-linear power spectrum is required at high significance to reconcile the \Planckns\ \LCDMns\ cosmology with DES (and KiDS) weak lensing measurements. 

%\medskip

\section{Small-scale power suppression: constraints on the parameter $A_{\rm \lowercase{mod}}$}
\label{sec:Amod}

Following \AE, in this section we incorporate the $A_{\rm mod}$ parameter to model suppression of power on non-linear scales via Eq.~\ref{equ:NL}. Variants 5 and 6 in Table~\ref{tab:chi2} use the full range of angular scales plotted in  Fig.~\ref{fig:OptSCvsNoSCvsAmodDV}. Variant 5 allows the cosmological parameters to vary freely over their priors, i.e. it is identical to variant 4 but with the addition of the $A_{\rm mod}$ parameter\footnote{\cite{Arico:2023} report constraints using DES weak lensing measurements without scale cuts \cite[see also][]{Chen:2022}. \cite{Arico:2023} use the baryonification prescription \citep{Schneider:2015,  Arico:2020} to model baryonic feedback but choose priors that restrict the strength of the feedback.} Variant 6 includes the \Planck prior of Eq.~\ref{equ:Planckprior}. 

\begin{figure*}
    \centering
    \begin{minipage}{.49\textwidth}
    	\includegraphics[width=0.93\columnwidth]{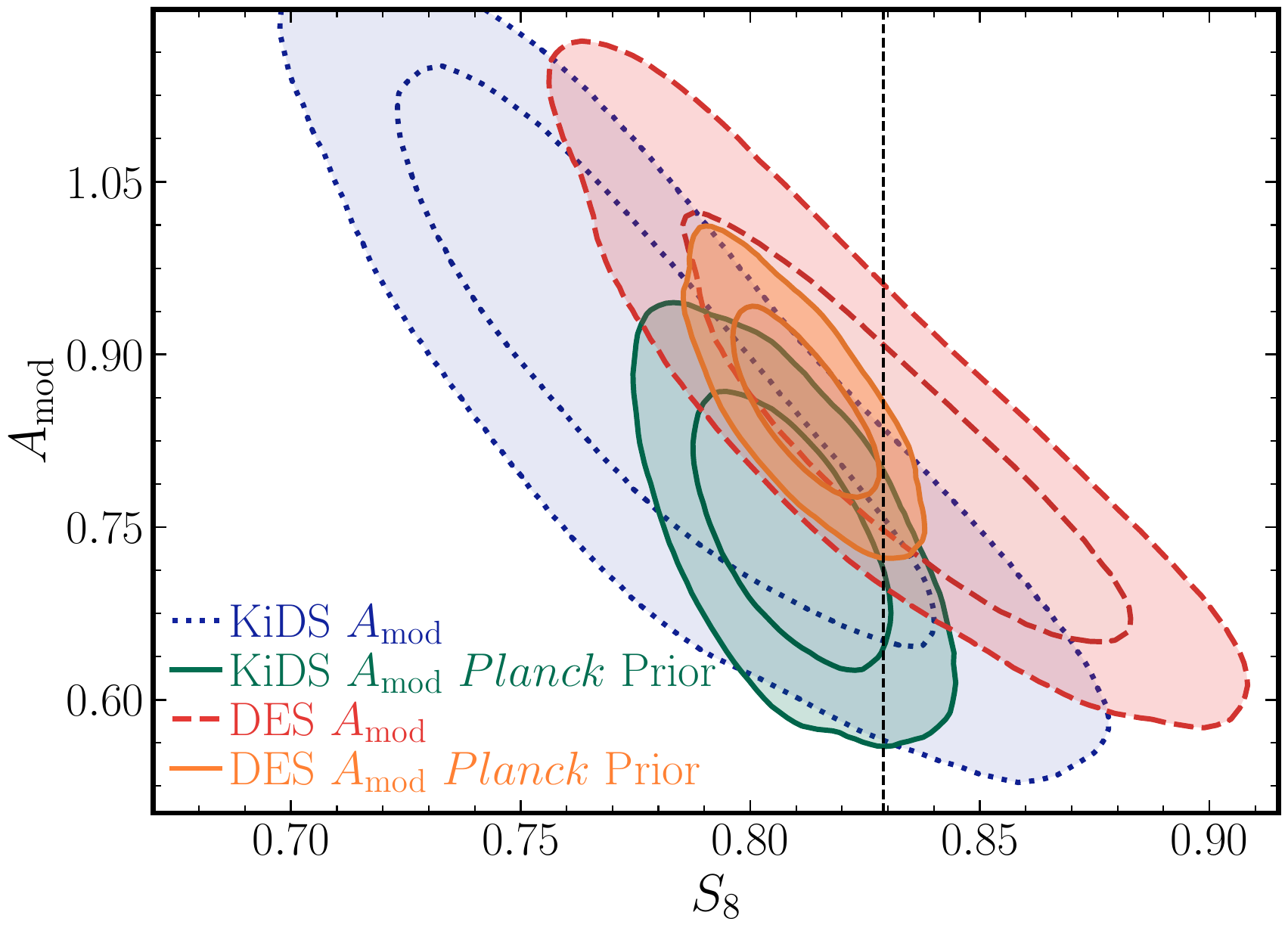} 
    \end{minipage}
    \begin{minipage}{.49\textwidth}
    \centering
    	\includegraphics[width=\columnwidth]{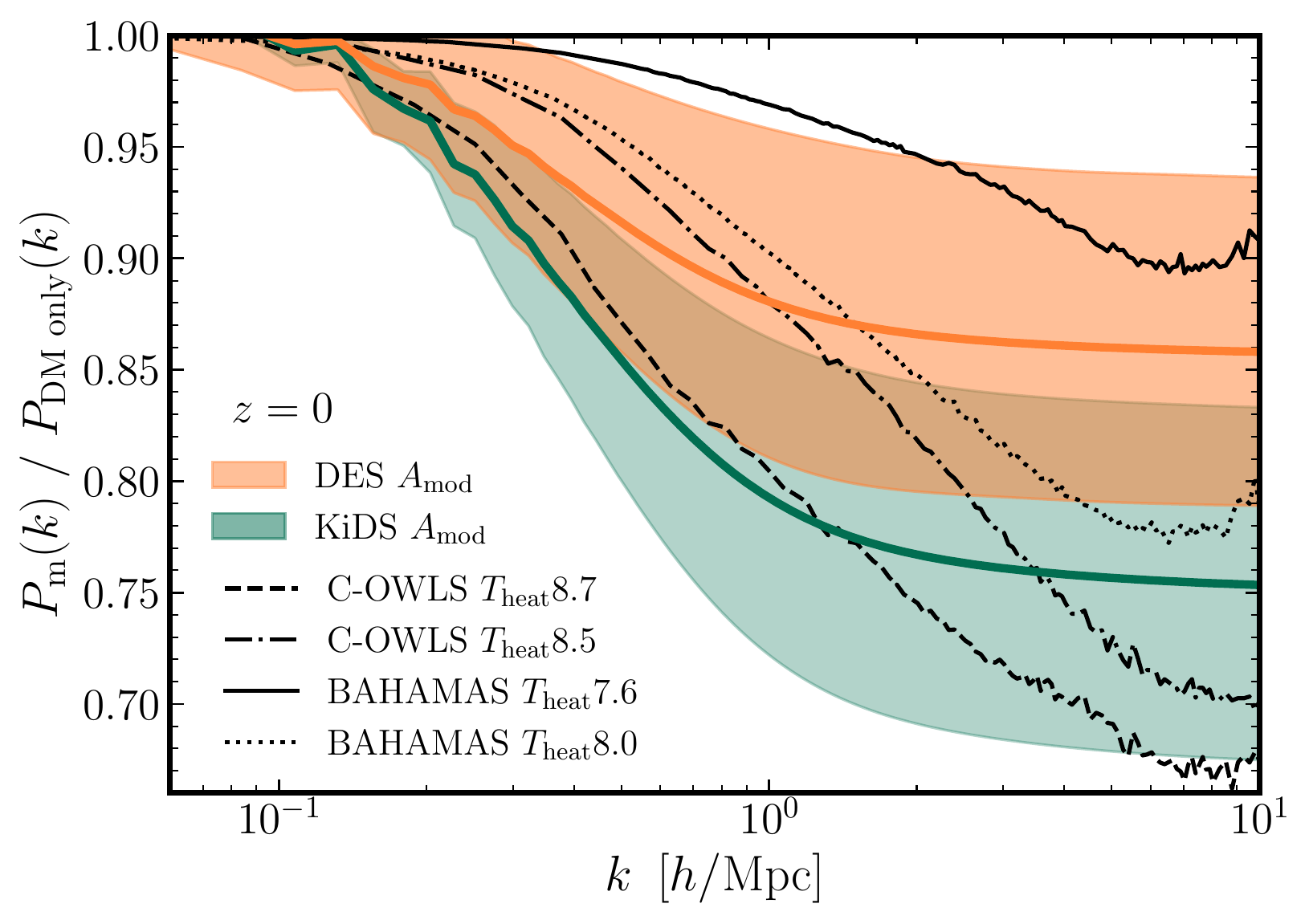} 
    \end{minipage}
\caption{Left: Illustration of the strong degeneracy between $S_8$ and the phenomenological power spectrum suppression parameter $A_{\rm mod}$ and the inferred  power suppression on non-linear scales. Left: the 68\% and 95\% constraints for DES $\xi_{\pm}$ statistics (without angular scale cuts) analyzed using {\sc HMCode2020}-no feedback, varying $A_{\rm mod}$ with free cosmology (red) and similarly for KiDS $\xi_{\pm}$ (blue), as presented in \AE. The dashed line indicates the \Planck \LCDM best-fit value of $S_{8}=0.828$. The DES constraint when incorporating the \Planckns-prior is shown in orange, and the KiDS equivalent in green. Right: the resulting power suppression corresponding to the DES/KiDS \Planckns-prior $A_{\rm mod}$ posteriors (orange/green). The DES data require less suppression of power compared to KiDS to be consistent with the \Planckns\ LCDM cosmology. As in \AE, we compare to predictions from various hydrodynamical simulations (black).}
\label{fig:AmodPk}
\end{figure*}

Fig.~\ref{fig:AmodPk} (left-hand side) compares the DES constraints from variants 5 (red) and 6 (yellow) in the $A_{\rm mod}-S_8$ plane. As shown in \AE\ and in Fig.~\ref{fig:AmodPk}, the $A_{\rm mod}$ parameter is strongly degenerate with the linear theory value of $S_8$. The vertical dashed line in Fig.~\ref{fig:AmodPk} shows the best fit \Planck\ \LCDMns\ value  $S_8 = 0.828$. The inclusion of the \Planck prior tightens the constraints significantly\footnote{Note that when we apply the $Planck$ prior, the nuisance parameters are fixed to their best-fit values. We have verified that freeing the nuisance parameters does not significantly alter the constraint on $A_{\rm mod}$. Similarly, the use of the $Planck$ prior does not cause any significant shifts in the nuisance parameters compared to the DES Y3 fiducial analysis. These tests show that the value of $A_{\rm mod}$ in Eq.~\ref{equ:amp} is insensitive to the nuisance parameters.}. One can see that adding the \Planck prior shifts the best fit value of $S_8$ downwards, but within about $1\sigma$ of the best fit \LCDMns\ value measured by \Planck\ alone. The \Planck \LCDM cosmology is therefore compatible with the DES Y3 weak lensing measurements provided that the power spectrum is suppressed on non-linear scales with 
\begin{equation}
      A_{\rm mod} = 0.858 \pm{0.052}, \qquad {\rm DES \ \ Y3 \ (no  \ scale  \ cuts}). \label{equ:amp} 
\end{equation}
Thus if no scale cuts are applied, $A_{\rm mod}$ differs from unity by about $2.7\sigma$. Note that the best fit model in variant 6 is plotted as the orange line in Fig.~\ref{fig:OptSCvsNoSCvsAmodDV} and is almost indistinguishable from the best fits to variants 3 and 4 which allow cosmological parameters to vary freely. 

Variant 7 is the same as variant 6 but applies the DES scale cuts. For this case,
\begin{equation}
      A_{\rm mod} = 0.919 \pm{0.099}, \qquad {\rm DES \ \ Y3 \ (scale  \ cuts}) \label{equ:amp_with_cuts}. 
\end{equation}
The error bar increases compared to Eq.~\ref{equ:amp} and now the parameter $A_{\rm mod}$ differs from unity by only about $0.8\sigma$. This, of course, does not conflict with Eq. \ref{equ:amp} and tells us merely that large angular scales are not as sensitive to power spectrum suppression as smaller angular scales. 

The choice of scale cuts in the DES Y3 analysis was based on the baryonic feedback effects measured from the EAGLE \citep{McAlpine:2016} and from the OWLS-AGN \citep{vanDaalen:2011} simulations, as detailed in \citet{Krause:2022}; \citet*{Secco:2022}. Using these simulations, scale cuts were determined which resulted in a maximum two-dimensional bias in the $\Omega_{\rm m}-S_8$ plane of $0.14\sigma_{2D}$ for the cosmic shear analysis. DES22 reported cosmological constraints by applying these scale cuts instead of modelling baryonic feedback effects.  Our results in Eq.~\ref{equ:amp_with_cuts} suggest that this strategy does largely reduce the sensitivity of cosmological results to power suppression on non-linear scales, since the value of $S_8$ in variant 3 is  within $0.8\sigma$ of the \Planckns\ \LCDMns\ value. However, as scale cuts are used to mitigate  baryonic effects, with no attempt to model feedback, the exact biases introduced into the $S_8$ parameter such will depend on the accuracy of the OWLS-AGN simulation as an upper bound on baryonic feedback effects. One cannot rule out small biases towards low values of $S_8$ if the baryonic feedback is actually stronger than in these simulations (or if the power spectrum suppression is caused by the properties of the dark matter). The results in Eq.~\ref{equ:amp_with_cuts} perhaps hint that this might be the case. What is clear, however, is that when the scale cuts are removed, significant suppression of the non-linear  spectrum is required to reconcile the DES Y3 lensing results with the \Planckns\ \LCDMns\ cosmology.

The values of $A_{\rm mod}$ in Eqs.~\ref{equ:amp} and \ref{equ:amp_with_cuts} are both higher than the value reported in \AE\ from an analysis of KiDS weak lensing. We compare KiDS with the result of Eq.~\ref{equ:amp}, where no scale cuts are used in either case (though we note that KiDS measurements extend to smaller angular scales than DES).
%The KiDS $\xi_\pm$ constraints make more aggressive use of small angular scales than the DES analysis with scale cuts and so we will compare KiDS with the result of Eq.~\ref{equ:amp}. 
In the analysis of KiDS reported in \AE, to infer a value of $A_{\rm mod}$ we kept the \LCDMns\ cosmological parameters fixed to the \Planck\ best fit values, whereas in this paper we have applied the \Planck\ prior of Eq.~\ref{equ:Planckprior}. Including a \Planck\ prior skews $A_{\rm mod}$ towards higher values  since the joint likelihood peaks at slightly lower values of $S_8$. We have repeated the analysis of \AE using $\xi_\pm$ including the \Planck\ prior (variant 13 in Table~\ref{tab:chi2}) finding
\begin{equation}
      A_{\rm mod} = 0.748 \pm{0.072}, \qquad {\rm KiDS} \label{equ:amp1},  
\end{equation}
which is $\sim 1\sigma$ higher than the value reported in AE. The joint constraints on $A_{\rm mod}$ and $S_8$ from this variant are shown by the filled blue contours in Fig.~\ref{fig:AmodPk}. The constraints from KiDS where cosmological parameters are allowed to vary freely (variant 12) are shown by the blue dotted contours. These constraints are similar to those from DES, though displaced to lower values of $S_8$. The differences between KiDS and DES are consistent with sampling fluctuations. Naively combining the two estimates Eqs.~\ref{equ:amp}-\ref{equ:amp1} we find
\begin{equation}
      A_{\rm mod} = 0.820 \pm{0.042}, \qquad {\rm DES \ \ Y3 + KiDS} \label{equ:amp2}.  
\end{equation}
Thus, according to our model, to reconcile the \Planckns\ base \LCDMns\ cosmology with DES+KiDS weak lensing data requires power spectrum suppression on small scales at high statistical significance ($\sim 4 \sigma$).

The power spectrum suppression corresponding to Eqs.~\ref{equ:amp} and~\ref{equ:amp1} at $z=0$ is shown in the right-hand panel of Fig.~\ref{fig:AmodPk}. In this and subsequent figures, we compare our predictions to the power spectrum suppression measured from hydrodynamical simulations taken from \cite{vandaalen:2020}. The solid and dotted lines correspond to the lower and upper bounds on the power spectrum suppression suggested by the BAHAMAS simulations \citep[BAryons and HAloes of MAssive Systems]{McCarthy:2017}. These correspond to their AGN feedback parameter of $\log_{10} ( \Delta T_{\rm heat}/K) = 7.6$ and $8.0$ respectively. With the feedback prescription adopted in BAHAMAS, the choice $\log_{10} ( \Delta T_{\rm heat}/K) = 7.8$ provides a good match to the observed gas fractions in groups and clusters and to the stellar mass function at $z = 0.1$ \citep{vandaalen:2020}. The feedback prescription used in the two  C-OWLS simulations \citep[COSMO-Overwhelmingly Large Simulations]{LeBrun:2014} is the same as in BAHAMAS and uses higher values of log$_{10}(\Delta T_{\rm heat}/K$). As a result, the power spectrum suppression in C-OWLS is stronger and extends to lower  wavenumbers than in the BAHAMAS simulations. As discussed in \citet{vandaalen:2020} the stronger feedback in these simulations resulted in deficit of galaxies with stellar masses below $10^{11} h^{-1} M_\odot$ compared to observations. For reference, the power spectrum suppression in the OWLS-AGN simulation used to select the DES angular scale cuts is similar to that in the fiducial BAHAMAS simulations with $\log_{10} (\Delta T_{\rm heat}/K) = 7.8$. The suppression in the EAGLE simulations is much weaker than any of the simulations plotted in Fig.~\ref{fig:AmodPk}, see \cite{Chisari:2019}.
 
To reconcile \Planckns\ \LCDMns\ with weak lensing based on DES and KiDS, the power spectrum on non-linear scales must be more strongly suppressed than in the BAHAMAS simulation with $\log_{10} ( \Delta T_{\rm heat}/K) = 8$  (which over the wavenumber range $k \simlt 1 h {\rm Mpc}^{-1}$ is well approximated by $A_{\rm mod} = 0.9$). Stronger power spectrum suppression, closer to the two C-OWLS models plotted in Fig.~\ref{fig:AmodPk} is required to match the results of Eqs.~\ref{equ:amp} and \ref{equ:amp1}. 

The one parameter $A_{\rm mod}$ model has the virtue of simplicity, but severely restricts the functional form of any power suppression. Ideally, one would want to reconstruct both the redshift and wavenumber dependence of the matter power spectrum directly from the observations. However, current weak lensing data are not sufficiently powerful to allow  a full 2-dimensional reconstruction. The next two sections should therefore be considered exploratory. Sec.~\ref{sec:Cmodz} investigates a simple parametric form for the redshift dependence, while Sec.~\ref{sec:abin} investigates the wavenumber dependence of the power spectrum  by splitting $A_{\rm mod}$ into a coarse set of bins in $k$.

\section{Redshift dependent power suppression}
\label{sec:Cmodz}

\begin{figure}
\centering
   \includegraphics[width=1\columnwidth]{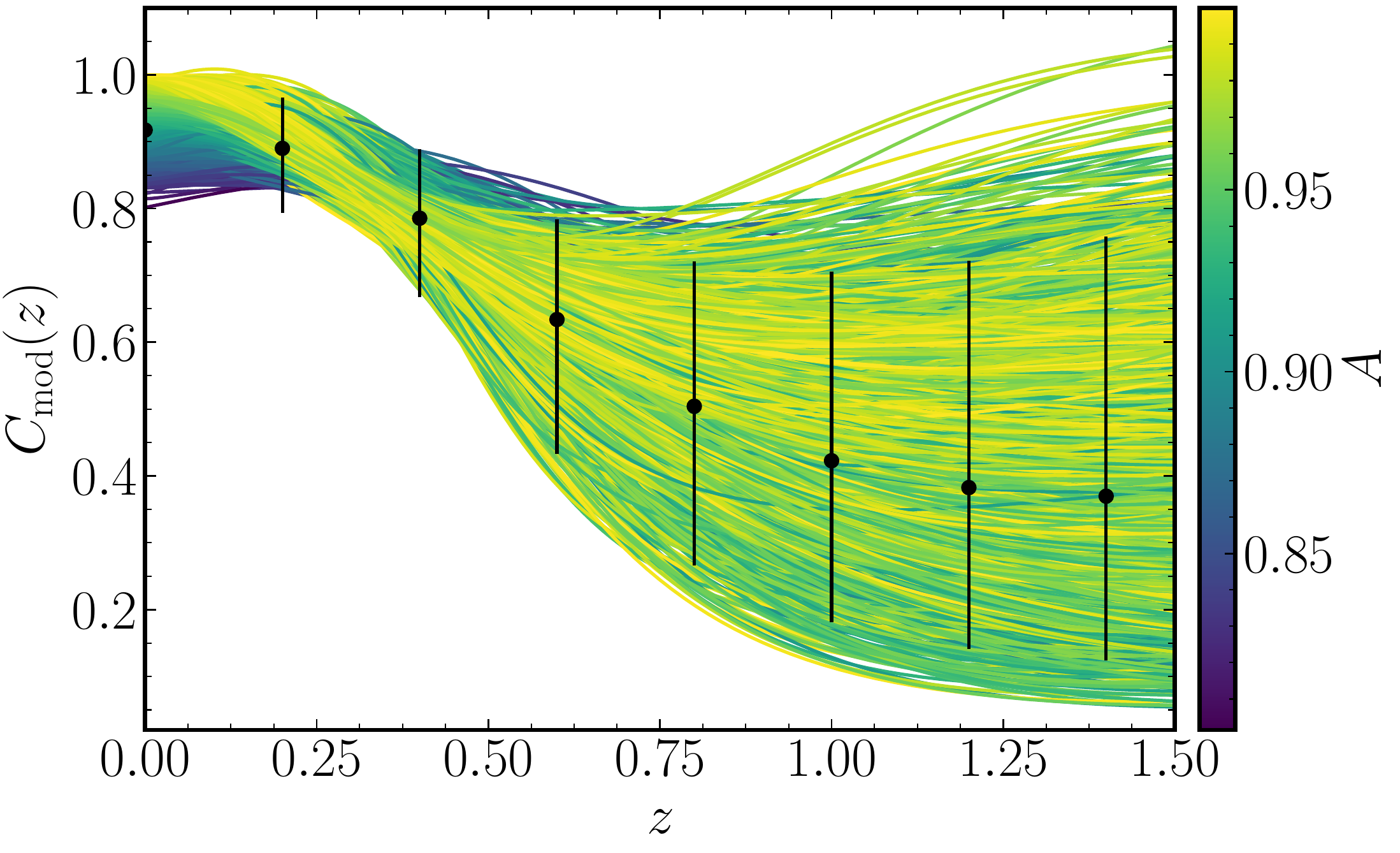}
\caption{Trajectories of $C_{\rm mod}(z)$ with $\Delta \chi^{2}< 1$ over the redshift range $z=0-1.5$. Overlaid black data points represent the weighted means of the trajectories, evaluated in intervals of $\Delta z=0.2$. The errors plotted here are at the 95\% confidence intervals and are asymmetric because the distributions are non-Gaussian
(as can be seen by the outliers). The trajectories are constrained most accurately  in the redshift region of $z \approx 0.3$; the redshift at which weak lensing surveys are most sensitive. At lower and higher redshifts the trajetories are not well constrained and depend sensitively on the choices of priors. The colour gradient indicates the corresponding value   $C_{\rm mod}(z = 0) = A$ with the parameterisation of Eq.~\ref{equ:NLZ}.}
\label{fig:CmodZEvolution}
\end{figure}
\begin{figure}
	\centering
	\includegraphics[width=\linewidth]{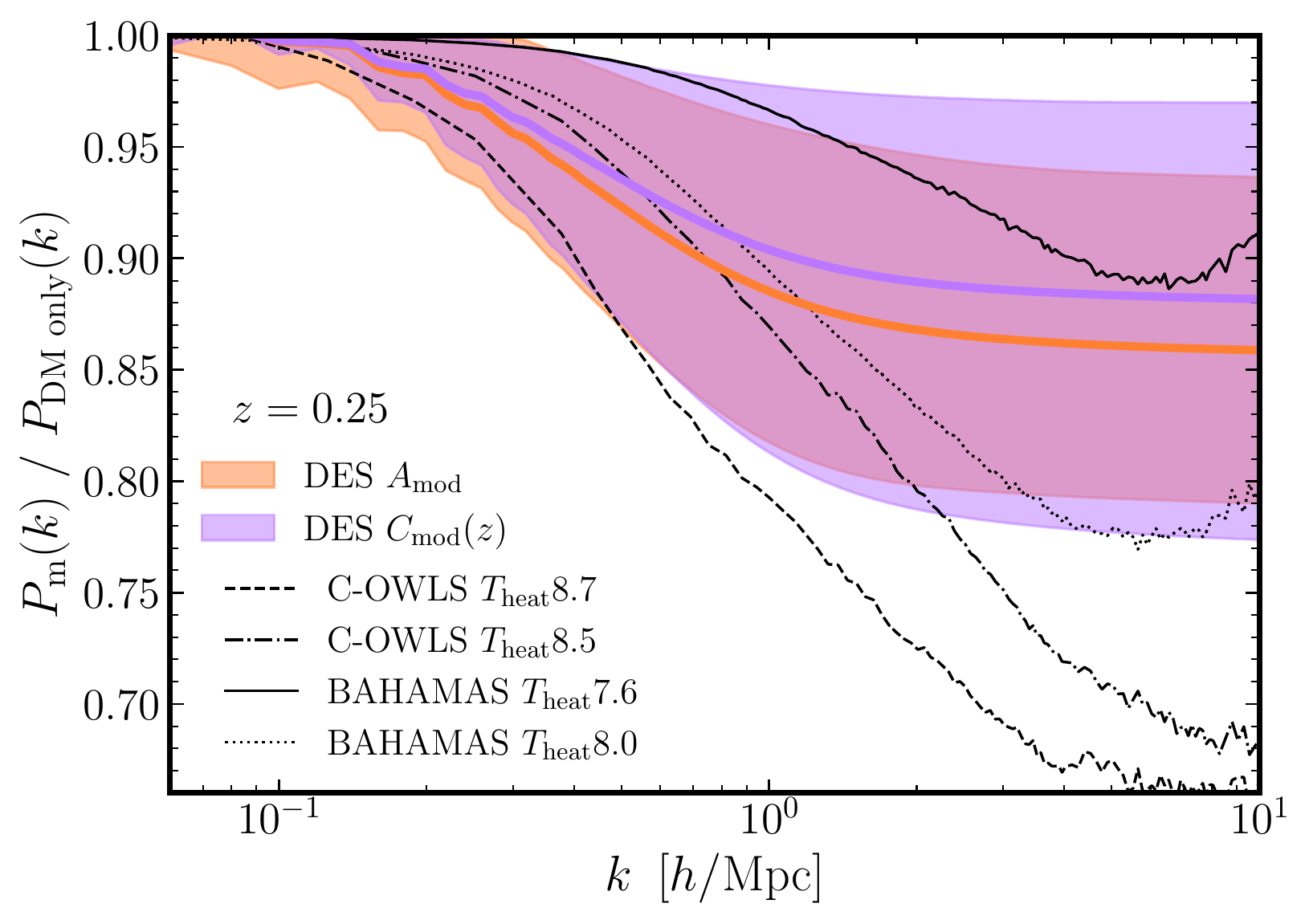} 
	\caption{\label{fig:Cmod} Power suppression predicted by the $C_{\rm mod}(z)$ model
   at a redshift $z=0.25$ (approximately the redshift at which DES weak lensing gives the tightest constraints) compared to the suppression seen in numerical simulations at
   this redshift.} 
\end{figure}

As discussed in DES22, the DES Y3 shear analysis is performed using four tomographic redshift bins with mean redshifts of $0.34$, $0.52$, $0.74$ and $0.96$. The mean redshift of the entire sample is $z_{\rm mean} \sim 0.63$ so we would expect the weak lensing statistics to be most sensitive to the matter distribution at a redshift $z\sim z_{\rm mean}/2\sim0.3$ and to have little sensitivity to the matter distribution at higher redshifts.

This reasoning can be made more quantitative by analysing  a parametric form of the redshift dependence of the matter power suppression. We chose the following functional form
\begin{subequations}
\begin{equation}
C_{\rm mod}(z) = A[1-f_{1}(z)] + B[f_{2}(z)],
\label{equ:CMODZ}
\end{equation} 
where the functions $f_k(z)$ are 
\begin{equation}
f_{k}(z) = \frac{(z/z_{\rm c_{k}})^{\alpha_{k}}}{1+(z/z_{\rm c_{k}})^{\alpha_{k}}}, \hspace{2mm}  k = 1,2. \label{equ:FZ}
\end{equation}
\end{subequations}
The matter power spectrum suppression according to this model is
\begin{equation}
    P_{\rm m}(k, z) =  P^{\rm L}_{\rm m}(k, z) + C_{\rm mod}(z)[P^{\rm NL}_{\rm m} (k, z) - P^{\rm L}_{\rm m}(k, z)],  \label{equ:NLZ}
\end{equation}
and uses six parameters $\{A$, $B$, $z_{c_1}$, $\alpha_1$, $z_{c_2}$, $\alpha_{2}\}$ instead of the single parameter $A_{\rm mod}$ of Eq.~\ref{equ:NL}. We adopt flat priors on these parameters as summarized in Table~\ref{tab:CmodZValues}.

\begin{table}
\setlength\extrarowheight{5pt}
  \centering
  \caption{Parameters of the $C_{\rm mod}(z)$ model of Sec.~\ref{sec:Cmodz}, prior ranges and their marginal mean values in both free and \Planck prior base \LCDM cosmologies. The priors on the parameters are uniform over the ranges given in the square brackets. Note that the redshift reconstructions are sensitive to the priors as noted in the text.}
\begin{tabular}{@{}l|c|cccc}
\hline
Parameter & Prior &  \multicolumn{2}{c}{Marginal Mean} &\\ \hline
 &  &  free cosmology & \Planck\ prior \\
 %  & \multicolumn{2}{|c|}{Display Format }& \\ \cline{3-4}
\hline

       \  $A$  & F[0.5, 1.0] & $0.893\pm{0.074}$ &  $0.914\pm{0.055}$  &    \\
       \  $B$  & F[0.5, 1.0]  & $0.759\pm{0.132}$ &  $0.757\pm{0.142}$ \\ 
       \  $z_{c_1}$ & F[0.1, 1.0]  & $0.674\pm{0.149}$ &  $0.768\pm{0.137}$   \\
       \  $z_{c_2}$ & F[1.0, 4.0] & $2.511\pm{0.817}$ & $2.472\pm{0.838}$ &  \\ 
       \  $\alpha_{1} $ & F[1.0, 4.0] & $2.472\pm{0.749}$ & $2.711\pm{0.728}$ \\
       \  $\alpha_{2} $ & F[1.0, 4.0]  & $2.491\pm{0.793}$ & $2.435\pm{0.860}$ \\ 

\hline
\end{tabular}
\label{tab:CmodZValues}
\end{table}

Figure~\ref{fig:Cmod} compares the power spectrum suppression constraints from variant 9 (purple contours) with those from the one-parameter $A_{\rm mod}$ model of variant 6 (orange contours). We show the results at a redshift of $z=0.25$,  
 since at this redshift the results are insensitive to our choice of the fitting function and priors.  The purple contours span a slightly broader range than the orange contours, since in the $C_{\rm mod}(z)$ model, one can trade off increased suppression at $z \sim 0.3$ with decreases in suppression at higher and lower redshifts and vice-versa. We also show
 the suppression measured in the cosmological hydrodynamics simulations at $z=0.25$. These are almost identical to the simulation results at $z=0$ shown in Fig.~\ref{fig:AmodPk}.

\begin{figure*}
	\centering
	\includegraphics[width=\textwidth]{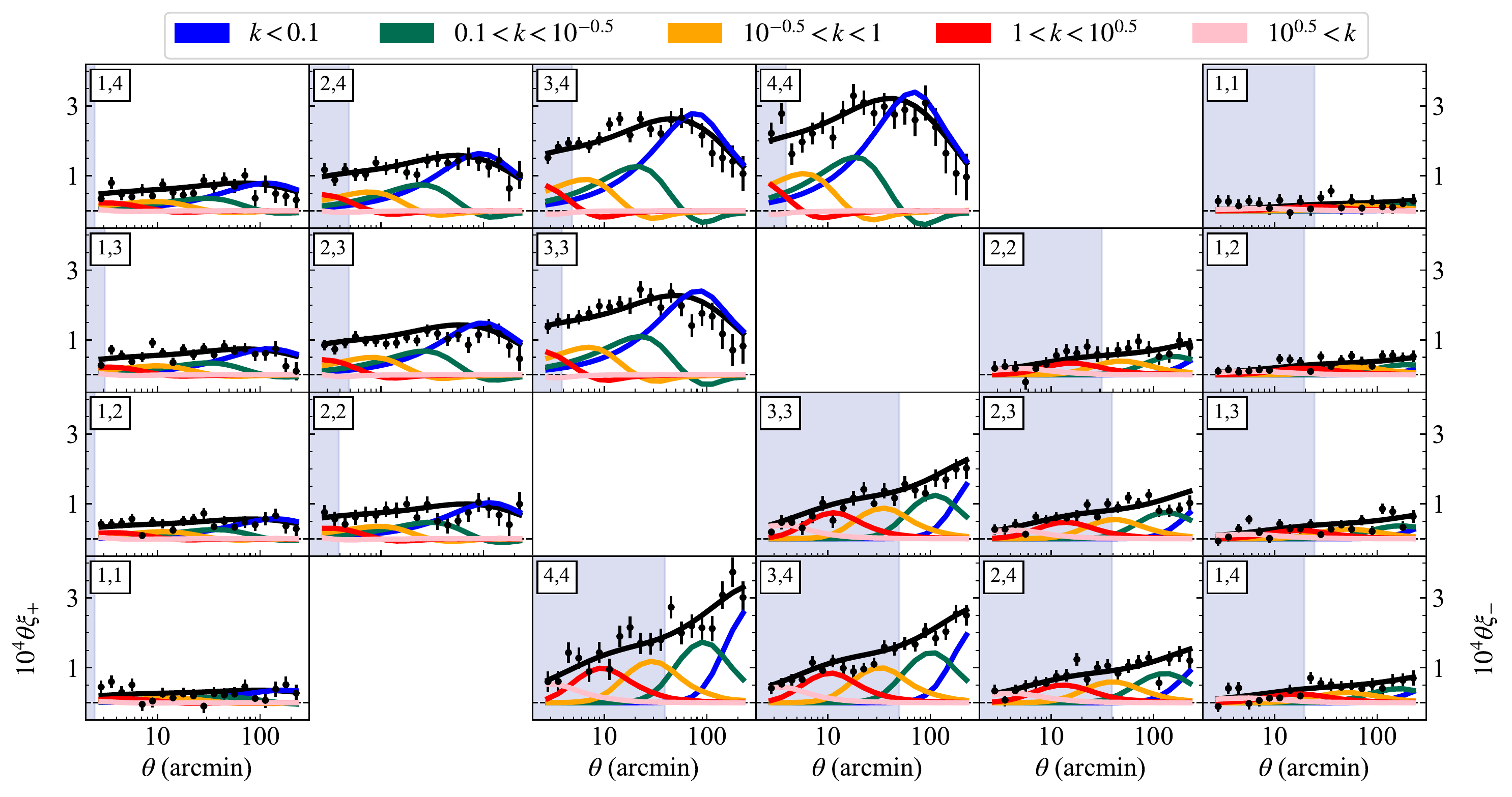}
	\caption{The best fit prediction for $\xi_{\pm}(\theta)$ (black) and the contributions of different $k$ ranges to the overall best fit are given in different colours, highlighting the relative power of each $k$-bin. The optimised \LCDM scale cuts from the DES Y3 analysis \citep{amon:2022}; \citep*{Secco:2022} are in shaded blue. The largest spatial scales (blue), corresponding to $k<0.1$, contribute to $\xi_{+}$ at large $\theta$ but make a very small contribution to $\xi_-$. The smallest scales (pink), $k>10^{0.5}$, contribute to $\xi_{-}$ on small angular scales but make a very small contribution to $\xi_{+}$.}
	\label{fig:DVScales}
\end{figure*}

In conclusion, the results of this section show that the DES data lack the statistical power to reconstruct the redshift dependence of the power suppression. Most of the statistical power of DES weak lensing comes from the matter distribution at $z \sim 0.3$, with relatively little sensitivity to the behaviour at higher or lower redshifts. The one parameter $A_{\rm mod}$ model cannot therefore be extrapolated beyond a relatively narrow range of redshifts centred at $z\sim 0.3$. 

\section{Scale-dependent suppression}
\label{sec:abin}

\begin{figure}
\centering
   \includegraphics[width=1\columnwidth]{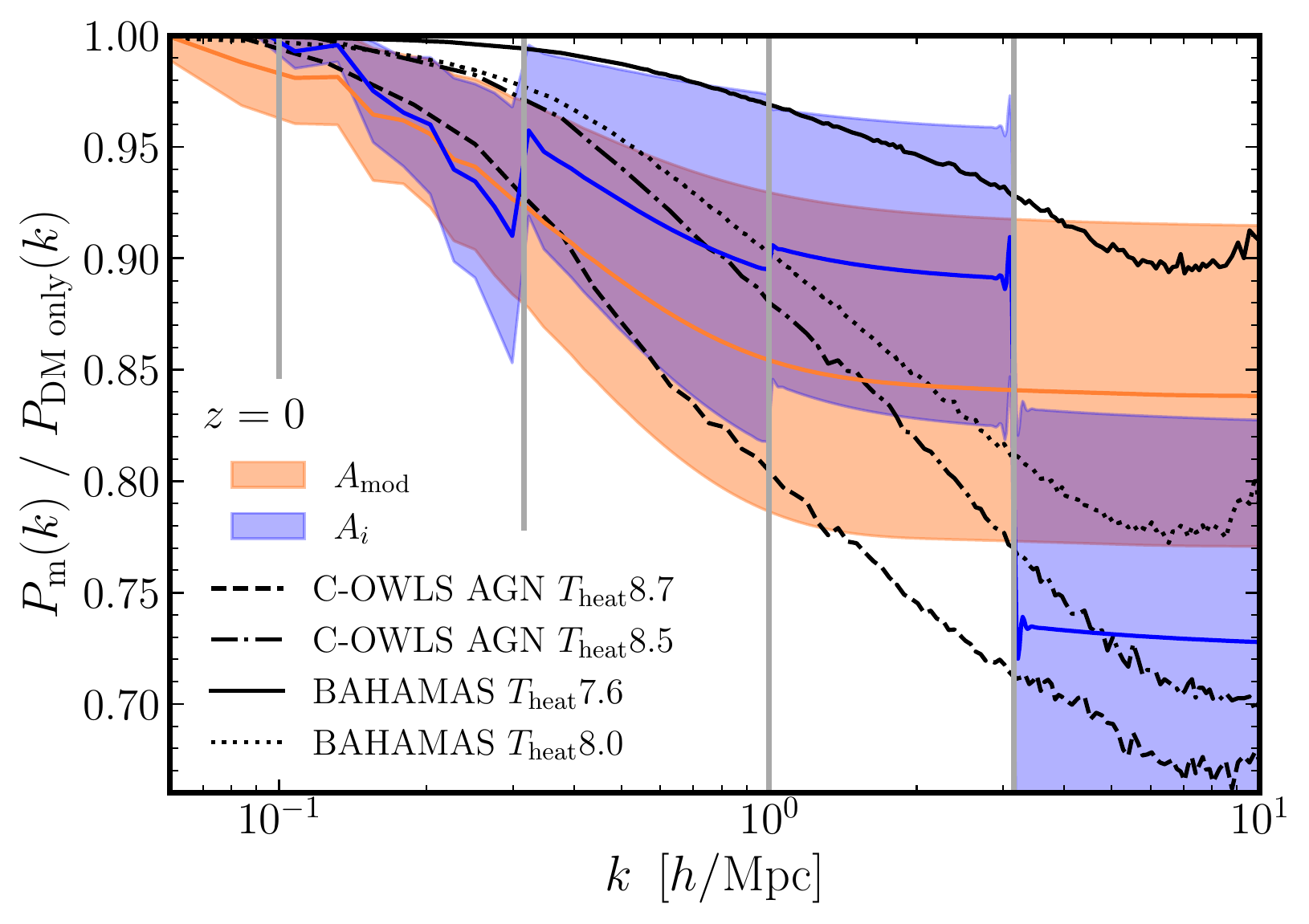}

\caption{The blue bands show the $1\sigma$ ranges of the reconstruction of the  power suppression using 5 wavenumber bins $A_i$ as in Eq. \ref{equ:Ai}. The constraints on the $A_{\rm mod}$ parameterisation from Fig. \ref{fig:AmodPk} are shown in orange. Note that the $A_i$  coefficients are strongly correlated with each other. We also plot the suppression measured in numerical hydrodynamicals simulations as in Fig. \ref{fig:AmodPk}.}
\label{fig:AB5Pk}
\end{figure}

In this section, we explore the scale-dependence of the non-linear matter power spectrum suppression by replacing the $A_{\rm mod}$ parameter with five parameters, $A_{i}$m separated in wavenumber $k$ as follows:
\begin{equation}
\left. 
\begin{array}{ll}
A_1: &  \ \ \ \ \ \ \ \ \ \ \log_{10} k  \le -1, \\
A_2: &  \ \ -1 < \log_{10} k \le 0.5, \\
A_3: &  -0.5 < \log_{10} k  \le 0, \\
A_4: &  \ \ \ \  0 < \log_{10} k  \le 0.5, \\
A_5: &  \ \ \ \ \ \ \ \ \ \   \log_{10} k  >  0.5. 
\end{array} \right \} \label{equ:Ai}
\end{equation}
where $k$ is in units of $h {\rm Mpc}^{-1}$ throughout this section.

To begin, we examine the contribution of each of the $k$-bins to the $\xi_{\pm}(\theta)$ prediction of the DES Y3 data\footnote{Specifically, we use the best-fit parameters from the analysis variant 4  in Table~\ref{tab:chi2}, with no scale cuts}, shown in Fig.~\ref{fig:DVScales}. We make several observations: (i) as is well known at any given angular scale $\theta$ the $\xi_{\pm}$ statistics mix of information from a wide range of $k$-scales;  (ii) $\xi_-$ is significantly more sensitive to non-linear information compared to $\xi_+$; (iii)  the bin covering linear scales ($k<0.1$, plotted in blue) makes little contribution to  $\xi_-$ over the measured angular range, though it is important for the larger $\theta$ values of $\xi_+$; (iv) the DES scale cuts effectively remove sensitivity to wavenumbers $k > 1$, thus small angular scales that are within the scale cuts must be included to constrain the parameters $A_4$ and $A_5$.

\begin{table}
\setlength\extrarowheight{5pt}
  \centering
  \caption{Parameters of the binned $A_{\rm mod}$ model. Prior ranges and marginal mean values in both free and fiducial \Planck base \LCDM cosmologies are given. Prior ranges are broad to allow for all possible trajectories to be investigated and are uniform (denoted by square brackets).}

\begin{tabular}{@{}l|c|cccc}
\hline

Parameter & Prior &  \multicolumn{2}{c}{Marginal Mean} &\\ \hline
 &  &  free cosmology & \Planck prior \\
\hline
     
       \  $A_{1}$ & F[0.5, 1.0] & $0.838\pm{0.190}$ &  $0.854\pm{0.200}$  &    \\
       \  $A_{2}$ & F[0.5, 1.0] & $0.913\pm{0.181}$ &  $0.815\pm{0.195}$ \\ 
       \  $A_{3}$ & F[0.5, 1.0] & $0.945\pm{0.138}$ &  $0.931\pm{0.117}$   \\
       \  $A_{4}$ & F[0.5, 1.0] & $0.894\pm{0.106}$ &  $0.903\pm{0.075}$ \\ 
       \  $A_{5}$ & F[0.5, 1.0] & $0.757\pm{0.117}$ &  $0.741\pm{0.104}$   \\

\hline
\end{tabular}
\label{tab:AB5Values}
\end{table}

We adopt flat priors on each $A_i$ parameter, as summarized in Table~\ref{tab:AB5Values}, with the lower limits describing much more extreme power suppression than seen in cosmological hydrodynamical simulations. The posterior mean value of $S_8$ and reduced-$\chi^{2}$ for the $A_{i}$ analysis are given in Table~\ref{tab:chi2} for both the free cosmology and \Planck prior cases (variants 10 and 11 respectively). The mean value of the posterior and the standard deviations of the $A_i$ parameters are reported in Table~\ref{tab:AB5Values}. 

Fig.~\ref{fig:AB5Pk} shows the predicted power suppression in each $k$-bin (separated by grey vertical lines), compared to the fiducial $A_{\rm mod}$ for the fits that include the \Planck prior. The constraints from the binned model track the general shape and amplitude of the one-parameter $A_{\rm mod}$ model. The main new result from this analysis is that power suppression of $\sim3-10$\%  spanning mildly non-linear scales (bin 2, spanning wavenumbers in the range $0.1 < k < 0.5$) is required to reconcile the \Planckns \LCDM\ data with the DES weak lensing data. It is not possible to avoid suppression in bin 2 by increasing the suppression at smaller scales, mainly because $\xi_{-}$ is dominated by bin 2 over the angular range $\theta \sim 40^\prime-100^\prime$ (see the green curves in Fig.~\ref{fig:DVScales}).

Fig.~\ref{fig:AB5Pk} also shows the power spectrum suppression measured in the BAHAMAS and C-OWLS simulations (as in Fig.~\ref{fig:AmodPk}). Evidently, if baryonic feedback is responsible for the apparent $S_8$ tension, the analysis of this section shows that the feedback must propagate to scales $k \simlt 0.3$. This requires stronger  feedback than in the  BAHAMAS simulation with $\log_{10} (\Delta T_{\rm AGN}/{\rm K}) = 7.8$  favoured by \citep{McCarthy:2017}, in agreement with the conclusions of Sec. \ref{sec:Amod}.  

\section{Discussion and Conclusion}
\label{sec:discussion}

\begin{figure}
	\centering
	\includegraphics[width=\columnwidth]{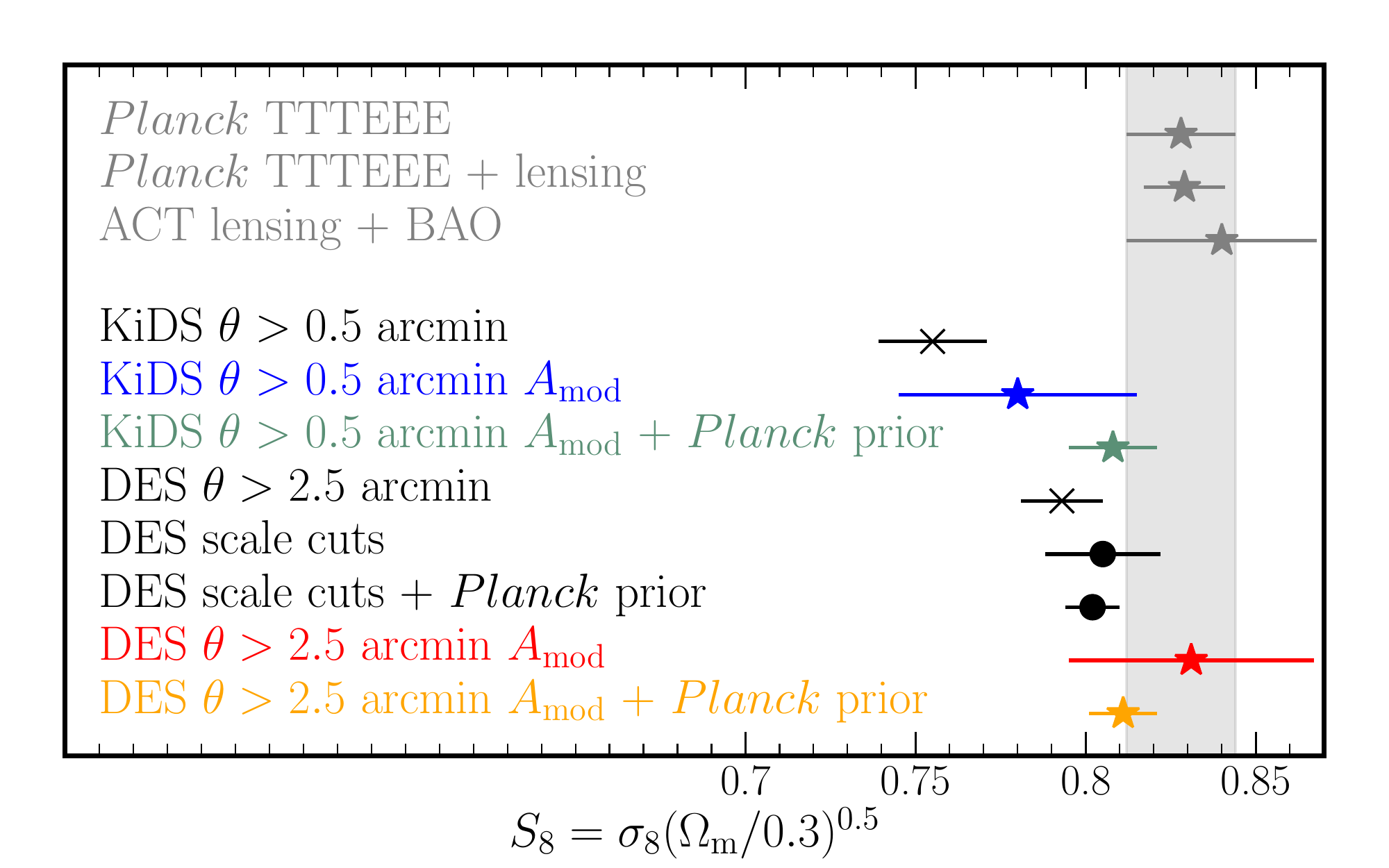}
	\caption{The \Planck TTTEEE, both with and without \Planck lensing \citep{EfstathiouGratton:2021} and the ACT+BAO constrains \citep{ACT_madhavacheril2023} $1\sigma$ constraints on $S_{8}$ are shown in grey. Below these, in black, blue and green, are the results for the KiDS data analysed with no scale cuts, no scale cuts with $A_{\rm mod}$ (from AAGE) and no scale cuts with both a \Planck prior and $A_{\rm mod}$ variants 12 and 13 from Table~\ref{tab:chi2} respectively. The next three results show in order the results of variants 4, 3 and adding a \Planck prior to the DES analysis of variant 3, respectively (black). Finally, the results of variants 5 and 6 are shown in red and yellow respectively. This illustrates the sensitivity of the $S_8$ tension on scale cuts and the modelling of non-linear scales.}
	\label{fig:S8summary}
\end{figure}

\begin{figure}
	\centering
	\includegraphics[width=\columnwidth]{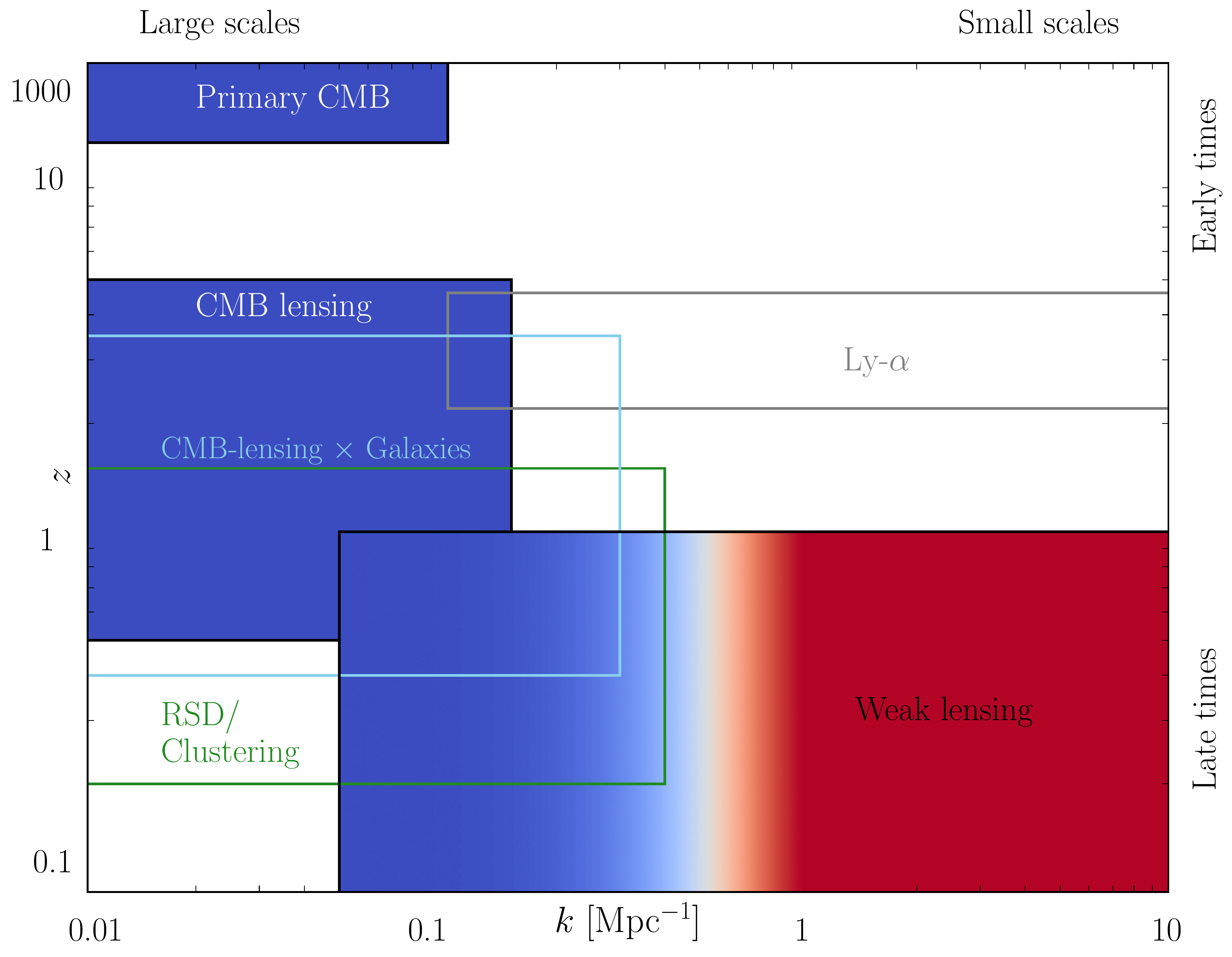}
	\caption{A rough guide to the approximate scale-dependence in terms of wavenumber, $k$, and redshift-dependence, $z$,  of cosmological observations. CMB lensing measurements are consistent with \Planckns\ \LCDMns\ (both blue, filled) and have negligible sensitivity on non-linear modelling and span the range $z\sim0.5-5$. Weak galaxy lensing is sensitive to a wide range of scales at $z<1$, but primarily probes the non-linear regime (red, filled). With future lensing data, is it possible to separate the linear information from weak-lensing. Both redshift space distortions and cross-correlations of CMB lensing with galaxy positions typically limit their analyses to linear scales and are sensitive to lower redshifts than CMB lensing. These two probes therefore provide a powerful test of the non-linear solution to the $S_8$ tension proposed here. Lyman-alpha measurements are also sensitive to a wide range of scales, but at higher redshifts.}
	\label{fig:KvsZprobes}
\end{figure}

The aim of this investigation has been to assess whether the $S_8$ tension can be resolved, that is \Planckns\ \LCDMns\ cosmology can be made consistent with weak lensing observations by modifying the matter power spectrum on non-linear scales. Following \AE, we have investigated constraints on the power suppression parameter, $A_{\rm mod}$ of Eq.~\ref{equ:NL}, using DES Y3 cosmic shear data. In this analysis we include a \Planck\ prior describing their constraints on key cosmological parameters and the associated uncertainties.

The DES data require substantial suppression of the matter power spectrum on non-linear scales to become consistent with \Planckns. The suppression required is less extreme than found from the KiDS weak lensing measurements, though the results from these two surveys are statistically consistent. However, if such a suppression is interpreted in terms of baryonic feedback, then it must be stronger than the most extreme feedback prescription implemented in the BAHAMAS simulations.

The constraints on $A_{\rm mod}$ depend on the angular scale cuts applied to the $\xi_{\pm}$ measurements. If the DES `\LCDMns-Optimised' angular scale cuts are imposed on $\xi_\pm$, the cosmological constraints from DES data are degraded and are statistically compatible with the \Planckns\ cosmology. For this case, $A_{\rm mod}$ is consistent with unity, though with a large error. 

We have analysed the DES Y3 data using an extended $A_{\rm mod}$ model that includes either a redshift or wavenumber dependence.
The DES data have little sensitivity to redshifts outside of a relatively narrow range centred at $z\sim 0.3$. The one parameter $A_{\rm mod}$ model, therefore, provides an adequate approximation at this redshift but cannot be extrapolated reliably to higher or lower redshifts. 

To investigate the wavenumber dependence, we solved for amplitude suppression factors $A_i$ in five logarithmically spaced bins. The results show that consistency between DES and \Planckns\ \LCDMns\ requires suppression on scales $k \simlt 0.3\ h/\rm{Mpc}$. This result is in agreement with our results for $A_{\rm mod}$ and shows that the requirement of the data for power suppression on these scales is not an artefact of the simple $A_{\rm mod}$ parameterisation.

Fig.~\ref{fig:S8summary} summarizes both the updated results of \AE\ and this paper. The entry labelled \Planckns\ TTTEEE shows the \Planckns\ base \LCDM\ constraints from \cite{EfstathiouGratton:2021} and the entry labelled \Planckns\ TTTEEE+lensing includes the \Planck\ lensing likelihood \citep{Plensing:2020}. ACT lensing+BAO shows the \LCDM\ constraints from the recent ACT CMB lensing results combined with baryon acoustic oscillation measurements \citep{ACT_madhavacheril2023}. These measurements are derived from predominantly linear scales ($k\simlt0.1\ h/\rm{Mpc}$) and demonstrate that the \LCDMns\ cosmology provides a consistent description of the matter fluctuations from the redshift of recombination to the low redshifts that dominate the CMB lensing signal, $z \sim 0.5-5$. 

To compare the results from the CMB with those from weak gravitational lensing, it is necessary to extract the {\it linear} amplitude of the matter fluctuations from statistics that are dominated by non-linear scales. This requires an accurate model of the dark matter power spectrum on non-linear scales, including modifications caused by baryonic feedback. Incorrect modelling of the non-linear spectrum can therefore lead to an apparent tension between weak lensing estimates of $S_8$ and those measured from the CMB. 

This is illustrated by the remaining entries in Fig.~\ref{fig:S8summary} which summarize results from the KiDS and DES Y3 weak lensing. One can see that there are varying degrees of tension with the CMB. The two entries labelled `KiDS $\theta> 0.5\ {\rm arcmin}$' (from \AE)  and `DES $\theta > 2.5 \ {\rm arcmin}$' (variant 4) use the full angular ranges of $\xi_{\pm}$ reported by the two surveys (i.e. no scale cuts). For these entries Fig.~\ref{fig:S8summary} shows results on $S_8$ allowing cosmological parameters to vary freely, using HMCode2020 to model the non-linear spectrum and ignoring baryonic feedback. The `tension'\footnote{The numbers quoted here are based on the simple expression $(S^1_8-S^2_8)/\sqrt{\sigma^2_{S^1_8}+\sigma^2_{S^2_8}}$.} with the \Planck\ TTTEEE+lensing entry is $\sim 3.7\sigma$ for KiDS and $\sim 2.1\sigma$ for DES. If we apply the DES \LCDM-Optimised scale cuts (variant 3) the tension with the \Planckns\ $S_8$ drops to $\sim 1 \sigma$. This suggests that the scale cuts have largely eliminated sensitivity to non-linear power spectrum suppression, at the expense of increasing the error on $S_8$,  leading to consistency with \Planckns. The fact that DES $S_8$ in this case is slightly low may be because the effects of non-linear power spectrum suppression have not been eliminated entirely. 

If we impose a \Planck\ prior and include the parameter $A_{\rm mod}$, the DES and KiDS lensing can be made compatible with \Planckns, with or without scale cuts. The question then is whether the small scale suppression required is physically reasonable. As summarized above, if it is caused by baryonic feedback, then the suppression must be stronger than favoured in recent cosmological hydrodynamic simulations. However, underestimating baryonic effects is not the only plausible interpretation since the suppression may reflect non-standard properties of the dark matter on non-linear scales \citep[see e.g.][]{Poulin:2022,Rogers:2023}.

If the weak-lensing $S_8$ tension is caused by suppression of power on non-linear scales, what would we expect  for other cosmological measurements? Fig.~\ref{fig:KvsZprobes} shows a sketch of the wavenumber and redshift ranges covered by various cosmological observations.  As noted in \AE, all measures of $S_8$ at low redshift from linear scales should agree with the \Planckns\ cosmology. Specifically, CMB lensing measurements combined with baryon acoustic oscillations, such as recent measurements from Advanced ACT are consistent with the \Planckns\ cosmology \citep{ACT_madhavacheril2023}. Cross-correlations of CMB lensing with galaxy surveys can provide an important test of  this hypothesis. Forthcoming analyses using the latest CMB lensing maps from ACT \citep{Qu:2023} and the South Pole Telescope \citep{Omori:2022} should agree with the predictions of the \Planck \LCDM model. 

As reviewed in \AE\, redshift-space distortions measured from linear scales from current galaxy redshift surveys are consistent with our interpretation,  but not to high precision. This situation should change decisively  in the near future with RSD measurements from DESI \citep[see e.g.][and references therein]{Schlegel:2022}. Measurements of $S_8$ from quasar Ly-alpha lines  from DESI uniquely give access to high $k$ and $z \simgt 1$ and offer an opportunity to potentially break the degeneracy between baryonic feedback and non-standard dark matter. Tests of a redshift-dependent growth, $\sigma_8(z)$ or $S_8(z)$ using measurements from from linear scales should be consistent with \LCDM \citep{White2022, DESY3LCDMEX,Garcia-Garcia:2021}\footnote{However, deviations from the \LCDM\ growth rate have been reported when linear measurements are combined with weak galaxy lensing measurements that are sensitive to non-linear scales \citep{Garcia-Garcia:2021, DESSPT2023}.}.

Weak galaxy lensing measurements offer a window to a wide range of scales as  indicated in Fig.~\ref{fig:KvsZprobes}. In the future, it may be possible to reconstruct the power spectrum as a function of wavenumber and redshift from weak lensing data. Assuming that systematic errors can be sufficiently mitigated, it may become possible to accurately test the \LCDM\ cosmology on linear scales and to reconstruct the matter power spectrum on small scales using weak galaxy lensing data alone. Such an analysis would establish unambiguously whether the $S_8$ tension is caused by a deviation from the \LCDM\ model at late times, or whether it is an apparent tension caused by physics on non-linear scales. 

\section*{Acknowledgements}
The authors would like to thank Joe Zuntz and Niall MacCrann for help in handling \textsc{cosmosis} \citep{Zuntz:2015}. CP would like to thank Jessie Muir and Sujeong Lee for helpful discussions on modified growth of structure, and Anthony Challinor for useful discussions on CMB lensing. Alexandra Amon is supported by a Kavli Fellowship. George Efstathiou is supported by an Leverhulme Trust Emeritus Fellowship. Calvin Preston is supported by a Science and Technology Facilities Council studentship.

This project has used public archival data from both the Dark Energy Survey and the Kilo Degree Survey. Funding for the DES Projects has been provided by the U.S. Department of Energy, the U.S. National Science Foundation, the Ministry of Science and Education of Spain, the Science and Technology FacilitiesCouncil of the United Kingdom, the Higher Education Funding Council for England, the National Center for Supercomputing Applications at the University of Illinois at Urbana-Champaign, the Kavli Institute of Cosmological Physics at the University of Chicago, the Center for Cosmology and Astro-Particle Physics at the Ohio State University, the Mitchell Institute for Fundamental Physics and Astronomy at Texas A\&M University, Financiadora de Estudos e Projetos, Funda{\c c}{\~a}o Carlos Chagas Filho de Amparo {\`a} Pesquisa do Estado do Rio de Janeiro, Conselho Nacional de Desenvolvimento Cient{\'i}fico e Tecnol{\'o}gico and the Minist{\'e}rio da Ci{\^e}ncia, Tecnologia e Inova{\c c}{\~a}o, the Deutsche Forschungsgemeinschaft, and the Collaborating Institutions in the Dark Energy Survey.
The Collaborating Institutions are Argonne National Laboratory, the University of California at Santa Cruz, the University of Cambridge, Centro de Investigaciones Energ{\'e}ticas, Medioambientales y Tecnol{\'o}gicas-Madrid, the University of Chicago, University College London, the DES-Brazil Consortium, the University of Edinburgh, the Eidgen{\"o}ssische Technische Hochschule (ETH) Z{\"u}rich,  Fermi National Accelerator Laboratory, the University of Illinois at Urbana-Champaign, the Institut de Ci{\`e}ncies de l'Espai (IEEC/CSIC), the Institut de F{\'i}sica d'Altes Energies, Lawrence Berkeley National Laboratory, the Ludwig-Maximilians Universit{\"a}t M{\"u}nchen and the associated Excellence Cluster Universe, the University of Michigan, the National Optical Astronomy Observatory, the University of Nottingham, The Ohio State University, the OzDES Membership Consortium, the University of Pennsylvania, the University of Portsmouth, SLAC National Accelerator Laboratory, Stanford University, the University of Sussex, and Texas A\&M University. Based in part on observations at Cerro Tololo Inter-American Observatory, National Optical Astronomy Observatory, which is operated by the Association of Universities for Research in Astronomy (AURA) under a cooperative agreement with the National Science Foundation. Based on observations made with ESO Telescopes at the La Silla Paranal Observatory under programme IDs 177.A-3016, 177.A-3017, 177.A-3018 and 179.A-2004, and on data products produced by the KiDS consortium. The KiDS production team acknowledges support from: Deutsche Forschungsgemeinschaft, ERC, NOVA and NWO-M grants; Target; the University of Padova, and the University Federico II (Naples). 

\section*{Data availability}
No new data were generated or analysed in support of this research.

\bibliographystyle{mnras} 
\bibliography{lensing}

\appendix
\section{\Planck \LCDM\ prior}
\label{sec:planckprior}

\begin{figure}
\centering
\begin{subfigure}[b]{0.49\textwidth}
   \includegraphics[width=1\linewidth]{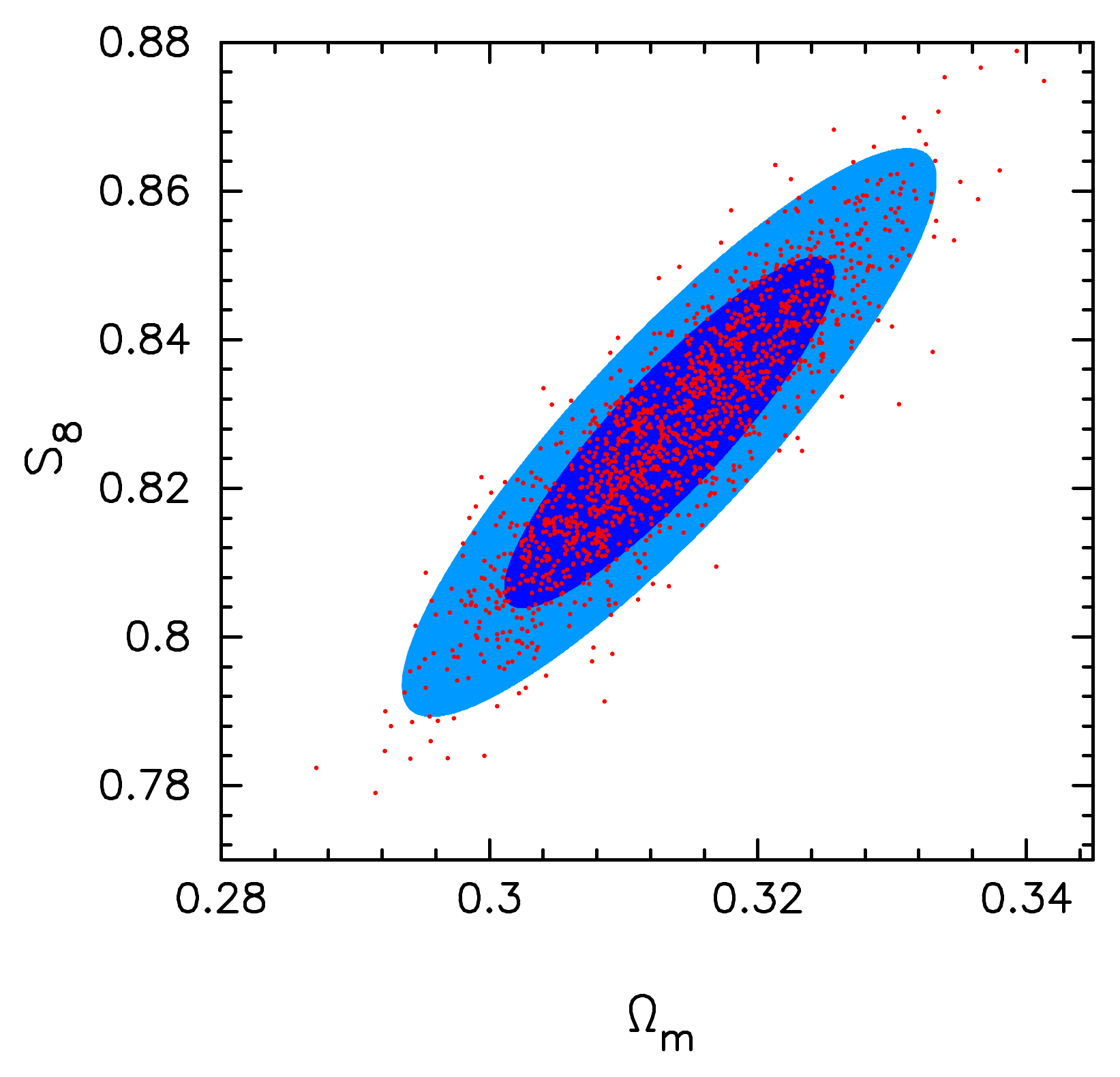}
   \label{fig:Ng1} 
\end{subfigure}

\begin{subfigure}[b]{0.49\textwidth}
   \includegraphics[width=1\linewidth]{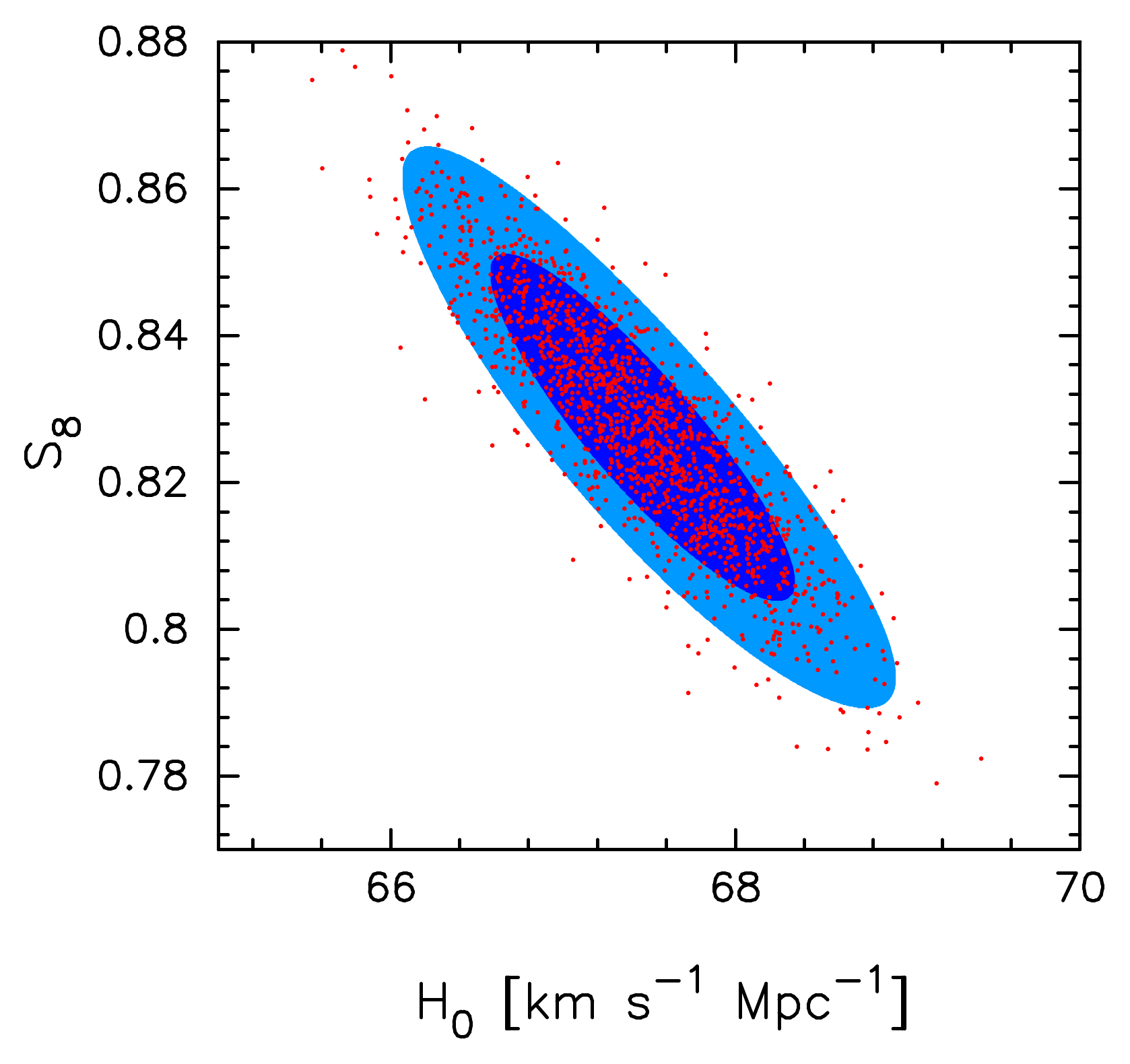}
   \label{fig:Ng2}
\end{subfigure}
\caption{\label{fig:planckprior} The degeneracies of $\Omega_{\rm m}$, $H_{0}$ and $S_{8}$ for the base \LCDM\ model using the \Planck TTTEEE likelihood computed by \citet{EfstathiouGratton:2021}. The ellipses show the 68\% and 95\% contours of $S_8$ and $\Omega_m$ (upper figure) and $S_8$ and $H_0$ (lower figure) computed from the two dimensional Gaussian model of Eq.~\ref{equ:Planckprior}. The contours in the lower panel assume Eq.~\ref{equ:Planckprior} and a delta function $\delta(\Omega_{\rm m} h^3- 0.09612)$. The parameter combination $\Omega_{\rm m} h^3$ is an approximation to the acoustic peak location parameter and is extremely well determined by \Planck\ ($\Omega_{\rm m} h^3 = 0.09612 \pm 0.00029$). The points show samples from the COSMOMC chains.}
\end{figure}

In this appendix we justify the \Planck\ prior of Eq.~\ref{equ:Planckprior}. In the base \LCDMns\ model, the spectral index $n_{\rm s}$ is well determined by \Planckns\ and can be fixed to the Planck\ best fit value. The main parameters that affect weak lensing are $S_8$, $\Omega_{\rm m}$ and $H_0$ (or combinations of these parameters). The parameter combination $\Omega_{\rm m} h^3$ is a proxy for the acoustic peak location parameter, $\theta_*$, and is extremely well determined by \Planckns. The \Planckns\ \LCDMns\ degeneracies in the space of $S_8$, $\Omega_{\rm m}$ and $H_0$ can be approximated accurately by adopting a two dimensional Gaussian in $S_8$ and $\Omega_{\rm m}$ and imposing a delta function constraint on the parameter combination $\Omega_{\rm m} h^3$. The specific form of Eq.~\ref{equ:Planckprior} comes from the inverse of the covariance matrix of the parameters $S_8$ and $\Omega_{\rm m}$ determined from the \textsc{COSMOMC} chains.  The degeneracies in the $S_8- \Omega_{\rm m}$ and $S_8-H_0$ planes inferred from Eq.~\ref{equ:Planckprior} are plotted in Fig.~\ref{fig:planckprior} and are compared to samples from the \textsc{COSMOMC} chains. 

\end{document}